\documentclass[twocolumn]{aastex7}

\usepackage{hyperref,enumerate}
\usepackage{xcolor}
\usepackage{amsmath}

\hypersetup{
    unicode=false,                  
    pdftoolbar=true,                
    pdfmenubar=true,                
    pdffitwindow=true,              
    pdfstartview={FitH},            
    pdftitle={MAGIC MIKE},          
    pdfauthor={vplacco},            
    pdfsubject={Astronomy},         
    pdfcreator={dvipdf},            
    pdfproducer={dvipdf},           
    pdfkeywords={metal-poor stars}, 
    pdfnewwindow=true,              
    colorlinks=true,                
    linkcolor=red,                  
    citecolor=blue,                 
    filecolor=magenta,              
    urlcolor=cyan,                  
    breaklinks=true,
    linktocpage
}

\usepackage{xcolor}

\definecolor{emerald}{rgb}{0.31, 0.78, 0.41}

\newcommand{\kmsec}{\mbox{km~s$^{\rm -1}$}}

\newcommand{\eps}[1]{\ensuremath{\log\epsilon\,(\mathrm{#1})}}
\newcommand{\vv}{{\tablenotemark{\footnotesize{a}}}}
\newcommand{\xx}{{\tablenotemark{\footnotesize{b}}}}
\newcommand{\yy}{{\tablenotemark{\footnotesize{c}}}}

\newcommand{\abund}[2]{\ensuremath{[\mathrm{#1}/\mathrm{#2}]}}
\newcommand{\cfe}{\abund{C}{Fe}}

\newcommand{\xfe}[1]{\abund{#1}{Fe}}

\newcommand{\metal}{\abund{Fe}{H}}

\newcommand{\teff}{\ensuremath{T_\mathrm{eff}}}
\newcommand{\logg}{\ensuremath{\log\,g}}
\newcommand{\Msun}{M$_\odot$}

\newcommand{\A}[1]{\ensuremath{A(\mathrm{#1})}}

\newcommand{\K}{\ensuremath{\mathrm{K}}}

\newcommand{\jone}{\object{\mbox{J0026$-$5445}}}
\newcommand{\jtwo}{\object{\mbox{J0433$-$5548}}}
\newcommand{\jthree}{\object{\mbox{J0711$-$5513}}}
\newcommand{\jfour}{\object{\mbox{J0712$-$5422}}}
\newcommand{\jfive}{\object{\mbox{J0717$-$6019}}}
\newcommand{\jsix}{\object{\mbox{J2336$-$5400}}}

\newcommand{\gui}[1]{{\textcolor{black}{#1}}}

\submitjournal{ApJ}

\begin{document}

\title{The DECam MAGIC Survey: Spectroscopic Follow-up of the \\ Most Metal-Poor Stars in the Distant Milky Way Halo\footnote{Based on observations gathered with the 6.5 m Magellan Telescopes located at Las Campanas Observatory, Chile.}}


\author[0000-0003-4479-1265]{Vinicius M.\ Placco}
\affiliation{NSF NOIRLab, Tucson, AZ 85719, USA}
\email{vinicius.placco@noirlab.edu}

\author[0000-0002-9269-8287]{Guilherme Limberg}
\affil{Kavli Institute for Cosmological Physics, University of Chicago, 5640 S. Ellis Avenue, Chicago, IL 60637, USA}
\email{}

\author[0000-0002-7155-679X]{Anirudh Chiti}
\affil{Department of Astronomy \& Astrophysics, University of Chicago, 5640 S. Ellis Avenue, Chicago, IL 60637, USA}
\affil{Kavli Institute for Cosmological Physics, University of Chicago, 5640 S. Ellis Avenue, Chicago, IL 60637, USA}
\email{}


\author[0000-0002-8217-5626]{Deepthi S. Prabhu}
\affil{Department of Astronomy/Steward Observatory, 933 North Cherry Avenue, Tucson, AZ 85721, USA}
\email{}

\author[0000-0002-4863-8842]{Alexander P. Ji}
\affil{Department of Astronomy \& Astrophysics, University of Chicago, 5640 S. Ellis Avenue, Chicago, IL 60637, USA}
\affil{Kavli Institute for Cosmological Physics, University of Chicago, 5640 S. Ellis Avenue, Chicago, IL 60637, USA}
\email{}


\author[0000-0002-8262-2246]{Fabr\'icia O. Barbosa}
\affiliation{Universidade de S\~ao Paulo, Instituto de Astronomia, Geof\'isica e Ci\^encias Atmosf\'ericas, Departamento de Astronomia, SP 05508-090, S\~ao Paulo, Brasil}
\email{}

\author[0000-0003-1697-7062]{William Cerny}
\affiliation{Department of Astronomy, Yale University, New Haven, CT 06520, USA}
\email{}

\author[0000-0002-6021-8760]{Andrew B. Pace}
\affiliation{Department of Astronomy, University of Virginia, 530 McCormick Road, Charlottesville, VA 22904, USA}
\email{}

\author[0000-0003-1479-3059]{Guy S. Stringfellow}
\affil{Center for Astrophysics and Space Astronomy, University of Colorado Boulder, Boulder, CO 80309, USA}
\email{}


\author[0000-0003-4102-380X]{David J. Sand}
\affil{Department of Astronomy/Steward Observatory, 933 North Cherry Avenue, Tucson, AZ 85721, USA}
\email{}

\author[0000-0002-9144-7726]{Clara E. Mart\'inez-V\'azquez}
\affil{NSF NOIRLab, Hilo, HI, 96720, USA}
\email{}

\author[0000-0001-5805-5766]{Alexander H. Riley}
\affil{Institute for Computational Cosmology, Department of Physics, Durham University, South Road, Durham DH1 3LE, UK}
\email{}

\author[0000-0001-7479-5756]{Silvia Rossi}
\affiliation{Universidade de S\~ao Paulo, Instituto de Astronomia, Geof\'isica e Ci\^encias Atmosf\'ericas, Departamento de Astronomia, SP 05508-090, S\~ao Paulo, Brasil}
\email{}

\author[0000-0002-8282-469X]{Noelia E. D. No\"el}
\affiliation{Department of Physics, University of Surrey, Guildford GU2 7XH, UK}
\email{}

\author[0000-0003-4341-6172]{A. Katherina Vivas}
\affiliation{Cerro Tololo Inter-American Observatory/NSF NOIRLab, Casilla 603, La Serena, Chile}
\email{}

\author[0000-0002-9269-8287]{Gustavo E. Medina}
\affiliation{Department of Astronomy and Astrophysics, University of Toronto, 50 St. George Street, Toronto, ON M5S 3H4, Canada}
\email{}


\author[0000-0001-8251-933X]{Alex Drlica-Wagner}
\affiliation{Fermi National Accelerator Laboratory, P.O.\ Box 500, Batavia, IL 60510, USA}
\affiliation{Kavli Institute for Cosmological Physics, University of Chicago, 5640 S. Ellis Avenue, Chicago, IL 60637, USA}
\affiliation{Department of Astronomy \& Astrophysics, University of Chicago, 5640 S. Ellis Avenue, Chicago, IL 60637, USA}
\affiliation{NSF-Simons AI Institute for the Sky (SkAI),172 E. Chestnut St., Chicago, IL 60611, USA}
\email{}

\author[0000-0002-1594-1466]{Joanna D. Sakowska}
\affiliation{Department of Physics, University of Surrey, Guildford GU2 7XH, UK}
\affiliation{Instituto de Astrofísica de Andalucía, CSIC, Glorieta de la Astronom\'\i a,  E-18080 Granada, Spain}
\email{}

\author[0000-0001-9649-4815]{Burçin Mutlu-Pakdil}
\affiliation{Department of Physics and Astronomy, Dartmouth College, Hanover, NH 03755, USA}
\email{}

\author[0000-0002-8093-7471]{Pol Massana}
\affiliation{NSF NOIRLab, Casilla 603, La Serena, Chile}
\email{}

\author[0000-0002-3690-105X]{Julio A. Carballo-Bello}
\affiliation{Instituto de Alta Investigaci\'on, Universidad de Tarapac\'a, Casilla 7D, Arica, Chile}
\email{}

\author[0000-0003-1680-1884]{Yumi Choi}
\affiliation{NSF NOIRLab, Tucson, AZ 85719, USA}
\email{}

\author[0000-0002-1763-4128]{Denija Crnojevi\'c}
\affiliation{Department of Physics \& Astronomy, University of Tampa, 401 West Kennedy Boulevard, Tampa, FL 33606, USA}
\email{}

\author[0000-0003-0478-0473]{Chin~Yi~Tan}
\affil{Department of Physics, University of Chicago, Chicago, IL 60637, USA}
\affil{Kavli Institute for Cosmological Physics, University of Chicago, 5640 S. Ellis Avenue, Chicago, IL 60637, USA}
\email{}


\collaboration{24}{(MAGIC \& DELVE Collaborations)}


\correspondingauthor{Vinicius M.\ Placco}

\begin{abstract}

In this work, we present high-resolution spectroscopic observations for six metal-poor stars with \metal$<-3$ (including one with \metal$<-4$), selected using narrow-band \ion{Ca}{2}~HK photometry from the DECam MAGIC Survey. The spectroscopic data confirms the accuracy of the photometric metallicities and allows for the determination of chemical abundances for 16 elements, from carbon to barium. The program stars have chemical abundances consistent with this metallicity range. A kinematic/dynamical analysis suggests that all program stars belong to the distant Milky Way halo population (heliocentric distances $35 < d_{\rm helio}/{\rm kpc} \lesssim 55$), including three with high-energy orbits that might have been associated with the Magellanic system and one, \jone, having parameters consistent with being a member of the Sagittarius stream. The remaining two stars show kinematics consistent with the Gaia-Sausage/Enceladus dwarf galaxy merger. \jtwo, with \metal=$-4.12$, is a carbon-enhanced ultra metal-poor star, with \cfe=$+1.73$. This star is believed to be a bona fide second-generation star, and its chemical abundance pattern was compared with yields from metal-free supernova models. Results suggest that \jtwo\, could have been formed from a gas cloud enriched by a single supernova explosion from a $\sim11$~\Msun\, star in the early universe. The successful identification of such objects demonstrates the reliability of photometric metallicity estimates, which can be used for target selection and statistical studies of faint targets in the Milky Way and its satellite population. These discoveries illustrate the power of measuring chemical abundances of metal-poor Milky Way halo stars to learn more about early galaxy formation and evolution.
\vspace{-0.75cm}
\end{abstract}


\section{Introduction} 
\label{intro}

The ever-changing chemistry of the universe is driven by stellar evolution. The current working model assumes that the first generation of stars was, in its majority, massive and metal-free, with a chemical composition reflecting that of the Big Bang \citep{oshea2008,bromm2013,klessen2023}. These stars, known as Population~III (hereafter Pop. III), were hypothesized almost 60 years ago by \citet{peebles1967}. However, the authors assumed that these stars would help solve the ``missing helium problem'' in young galaxies rather than drive the enrichment of the interstellar medium in heavier elements. Shortly after, \citet{ezer1971} built models for the evolution of hydrogen-helium stars under the assumption that the stars responsible for the early metal enrichment in galaxies must all be massive \citep{schmidt1963,truran1971}. 

The connection between Pop.~III and Galactic metal-poor stars was first posed by \citet{wallerstein1963} and later by \citet{truran1971mp}, arguing that these second-generation stars originated when the interstellar medium was almost entirely hydrogen, and their metallic contents were formed and ejected from massive, rapidly evolving metal-free stars. These long-lived, low-mass ``descendants of the first stars''  are metal-poor\footnote{To our knowledge, the first recorded mention to ``metal-poor stars'' was given by \citet{greenstein1958}, who determined the abundances of metals, CN, and CH in giant stars. The metallicities for their sample stars were slightly sub-solar since many of their targets were selected as carbon stars, based on the earlier work of \citet{keenan1942}, \citet{bidelman1956}.} and retain the record of interstellar gas abundances before much of the nucleosynthesis occurred in the early universe. 
Since then, it became clear that the so-called Extremely and Ultra Metal-Poor \citep[EMP and UMP -- \metal\footnote{\abund{A}{B} = $\log(N_A/{}N_B)_{\star} - \log(N_A/{}N_B) _{\odot}$, where $N$ is the number density of atoms of a given element in the star ($\star$) and the Sun ($\odot$), respectively.}$\leq-3.0$ and $\leq-4.0$, respectively;][]{frebel2015} stars hold in their atmospheres key information about the nature, distribution, and evolution of Pop.~III stars.

Besides having low iron content, \citet{rossi1999} recognized that, as the metallicity decreases in samples of field metal-poor stars, the fraction of objects showing enhancements in carbon increases, reaching 80\% for \metal$\leq-4.0$ and 100\% for \metal$\leq-5.0$ \citep{lee2013,placco2014c,arentsen2022}. These are known as Carbon-Enhanced Metal-Poor stars \citep[CEMP -- \cfe$\geq+0.7$;][]{aoki2007} and have been the subject of extensive studies over the past three decades\footnote{It is worth pointing out that the correlation between metal deficiency and carbon enhancement had been recognized qualitatively much earlier by \citet{hale1898} and \citet{duner1899}.}. The elevated carbon (coupled with nitrogen and oxygen) and the light-element ($Z\leq30$) chemical abundance pattern of CEMP stars in the UMP regime have been shown to closely resemble those from a variety of theoretical models of metal-free massive star explosions \citep[][among others]{heger2002,heger2010,lee2024,meynet2010,nomoto1999,nomoto2006,nomoto2013,tominaga2007,umeda2005}. Additional compelling evidence connecting Galactic metal-poor stars with the early universe resides on Damped Ly$\alpha$ systems at high-redshift, which show light-element abundance patterns that qualitatively resemble the ones of EMP and UMP stars \citep{cooke2011,cooke2012,cooke2014,banados2019} and also provide good fits for metal-free Pop.~III models.


Despite extensive targeted searches, only about 50 stars are known to have \metal$\leq-4.0$ from high-resolution ($R\geq20,000$) spectroscopy \citep{bonifacio2025}. One of the main early contributors has been the Hamburg/ESO Survey \citep[HES;][]{reimers1997,christliebhes01,christlieb2008}, which identified the first two stars in the Galactic halo with \metal$\leq-5.0$ \citep{christlieb2002,frebel2005}. Most confirmed EMP and UMP from HES have first been vetted by medium-resolution ($R\sim2,000$) spectroscopic follow-up \citep{christlieb2001,frebel2006,placco2010,placco2011}, which was a required step to determine metallicities before gathering high-resolution data. 

There are, however, efforts to identify EMP and UMP candidates directly from photometry, using a narrow-band filter centered in the metallicity-sensitive \ion{Ca}{2} H (3968.47\AA) and K (3933.66\AA) lines. The narrow-band magnitudes can be used to estimate photometric metallicities or in a color-color diagram to locate regions with a higher fraction of low-metallicity stars. Major efforts in this area include the Pristine Survey \citep{starkenburg2017}, the Javalambre Photometric Local Universe Survey \citep[J-PLUS;][]{cenarro2019,whitten2019,galarza2022}, the Southern Photometric Local Universe Survey \citep[S-PLUS;][]{mendesdeoliveira2019,whitten2021, splusSHORTSdr1}, and the Javalambre-Physics of the Accelerating Universe Astrophysical Survey \citep[J-PAS;][]{bonoli2021,yuan2023}. We refer the reader to \citet{placco2022} for a review of recent results provided by these surveys.  

The most recent effort to provide photometric \metal\, for 5,500 square degrees of the Milky Way and its structures in the Southern Sky is the Mapping the Ancient Galaxy in CaHK survey (MAGIC; NOIRLab Prop. ID 2023B-646244; P.I. Anirudh Chiti), a 54-night project conducted using the narrow-band \ion{Ca}{2}\,HK N395 filter installed on the Dark Energy Camera \citep[DECam;][]{flaugher2015} at the 4-m NSF Víctor M. Blanco Telescope on Cerro Tololo, Chile. The main goal is to derive photometric metallicities for stars down to the EMP and UMP regime by combining the new narrow-band photometry with existing DECam $g$ and $i$ coverage. MAGIC is organized within the DECam Local Volume Exploration Survey \citep[DELVE; NOIRLab Prop. ID 2019A-0305;][]{drlica2021,drlica2022} and further details about the survey strategy and early science will be part of an upcoming paper (Chiti et al., in preparation). An initial validation and application of MAGIC metallicities to the Sculptor dwarf galaxy is described in \citet{barbosa+2025}.

This article presents the discovery of five EMP stars and one UMP star in the outer Milky Way halo. This is the first high-resolution spectroscopic follow-up campaign from the MAGIC Survey and confirms the efficacy of selecting EMP star candidates from narrow-band photometry. The spectra, combined with data from the Gaia space mission \citep{gaia2016}, allowed for the determination of chemical abundances for up to 16 elements in each star, as well as an analysis of the dynamics and kinematics of the program stars. To test the efficacy of MAGIC targeting in characterizing new low metallicity populations, we specifically followed up stars in the distant Milky Way halo (with distances $d \gtrsim 30$\,kpc from the Sun) aiming to uncover the lowest metallicity stars in this Galactic component, trace the signatures of its early enrichment, and possibly associate any of these stars with outer halo substructures.

This work is outlined as follows: Section~\ref{observations} describes the target selection and spectroscopic observations, followed by the determination of stellar atmospheric parameters and chemical abundances in Section~\ref{atmparsec}. In Section~\ref{chemod}, we assess the quality of the photometric metallicities, compare the chemistry and kinematics of the program stars with data from the literature, and speculate on the origin of the UMP star in the sample, \jtwo. Conclusions and perspectives for future work are given in Section~\ref{conclusion}.

\begin{deluxetable*}{@{}cccccccc@{}}[!ht] 
\tabletypesize{\tiny}
\tabletypesize{\footnotesize}
\tablewidth{0pc}
\tablecaption{\label{obsinfo} Basic information for the program stars.}
\tablehead{
\colhead{Full ID}&
\colhead{Short ID}&
\colhead{Gaia DR3 ID}&
\colhead{Right Ascension}&
\colhead{Declination}&
\colhead{$g$}&
\colhead{\metal$_{\rm MAGIC}$}&
\colhead{Exp. (s)}}
\startdata
MAGIC~J002603.16$-$544500.3 & \jone   & 4923149526247840640 & 00:26:03.16 & $-$54:45:00.3 & 16.73 & $-$3.16 & 3000 \\
MAGIC~J043330.99$-$554839.2 & \jtwo   & 4775927771146196864 & 04:33:30.99 & $-$55:48:39.2 & 17.72 & $-$3.36 & 6900 \\
MAGIC~J071131.56$-$551326.5 & \jthree & 5490389769744376960 & 07:11:31.56 & $-$55:13:26.5 & 18.06 & $-$3.40 & 5100 \\
MAGIC~J071214.95$-$542227.7 & \jfour  & 5491209253801516288 & 07:12:14.95 & $-$54:22:27.7 & 17.33 & $-$3.28 & 2000 \\
MAGIC~J071706.29$-$601908.7 & \jfive  & 5293243525250093312 & 07:17:06.29 & $-$60:19:08.7 & 17.94 & $-$3.29 & 3600 \\
MAGIC~J233646.73$-$540022.1 & \jsix   & 6498000990151085696 & 23:36:46.73 & $-$54:00:22.1 & 16.86 & $-$3.09 & 4200 \\
\enddata
\end{deluxetable*}

\section{Target Selection and Observations}
\label{observations}

The six program stars studied in this work were selected as potential EMP outer halo star candidates in the MAGIC catalog\footnote{MAGIC internal collaboration catalog version \texttt{v250130}.} containing $\sim21$ million stellar sources from the first year of survey observations. The photometric metallicities (\metal$_{\rm photo}$) are calculated by comparing the observed narrow-band CaHK and broad-band DELVE DR2 $g$ and $i$ magnitudes to a grid of synthetic photometry, generated following the prescriptions in \citet{chiti2020, chiti2021a}, with the narrow-band CaHK filter replacing the SkyMapper $v$ filter \citep{bessel2011} in that work. An initial implementation of these photometric metallicities to data in the Sculptor dwarf galaxy is described in \citet{barbosa+2025}, which also outlines the methodology. Besides selecting sources with \metal$_{\rm photo}\leq -3.0$, we also restricted our sample to include only those having $g\lesssim18$ to allow for high-resolution follow-up spectroscopic observations within reasonable exposure times. A further selection of $g-i > 0.6$, Gaia DR3 \texttt{parallax/parallax\_error} $< 3.0$, and proper motion $< 5$\,mas\,yr$^{-1}$ was made to select cool red giants. The $g-i$ color of these stars was then matched to a 12\,Gyr, [Fe/H] = $-2.5$ Dartmouth isochrone to derive distance moduli for each target. Those with inferred distances $>30$\,kpc were then selected for observations\footnote{A total of 18 stars met this criteria, were observable from the southern hemisphere within the right ascension window, and also passed additional cuts, mostly avoiding the center of the Large and Small Magellanic Clouds.} to specifically target distant red giants in the outer Milky Way halo, where chemical abundances of the EMP/UMP field stellar population remain sparsely sampled \citep[e.g.,][]{battaglia2017,hansen2020}. 

\begin{figure}[!ht]
 \includegraphics[width=1\linewidth]{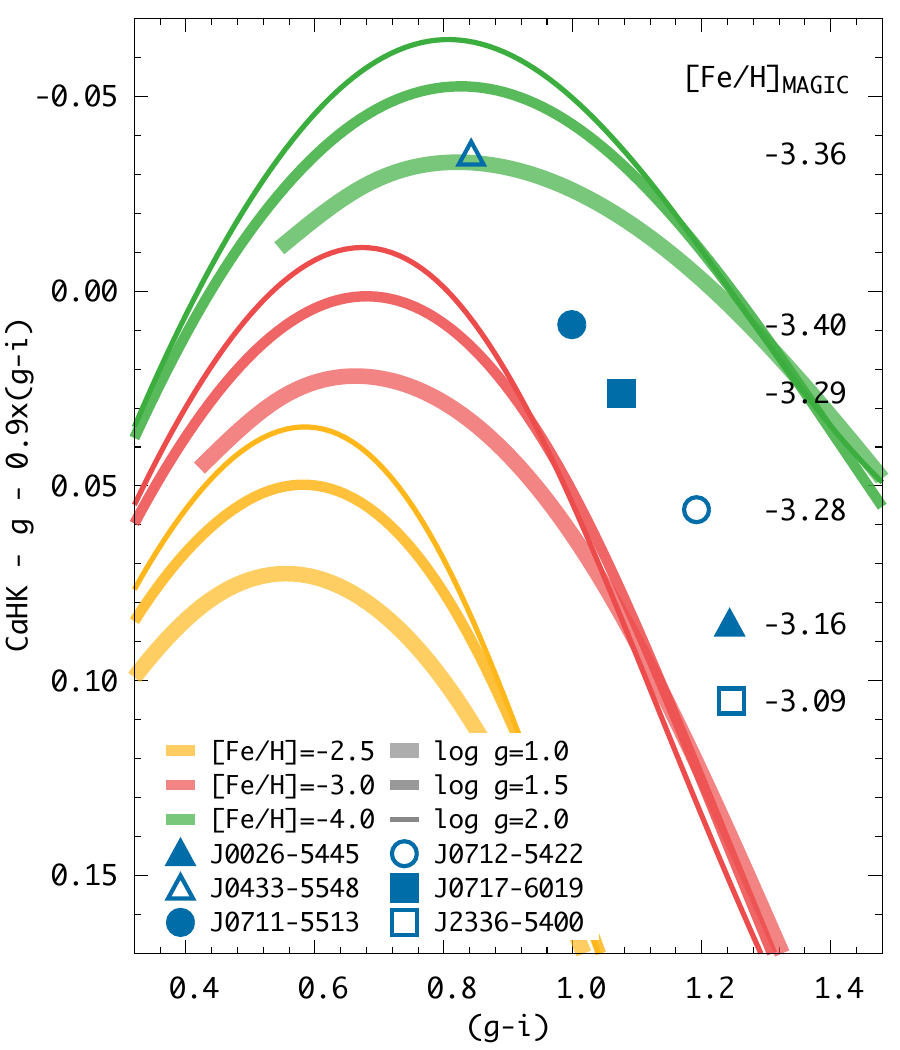}
     \caption{Color-color diagram for the program stars using the metallicity-sensitive CaHK magnitude. Colored lines represent constant \metal\, values, and the line thickness represents different \logg. The program stars are represented by symbols, with the photometric \metal\, values shown for reference.}
 \label{isofeh}
\end{figure}

Figure~\ref{isofeh} shows the metallicity-sensitive color-color diagram for the program stars (symbols). The colored solid lines represent constant metallicity for three different \logg\, values (line thickness) and were calculated by convolving a grid of synthetic spectra with filter transmission curves \citep[see][for further details]{chiti2021a}. Note that the program stars are located in the region embedded within the \metal=$-3.0$ and \metal=$-4.0$ tracks. The photometric metallicity values (\metal$_{\rm MAGIC}$) for each star are also shown for reference. In Section~\ref{photo}, we present a comparison between the photometric and spectroscopic \metal\, values.

\begin{figure*}[!ht]
 \includegraphics[width=1\linewidth]{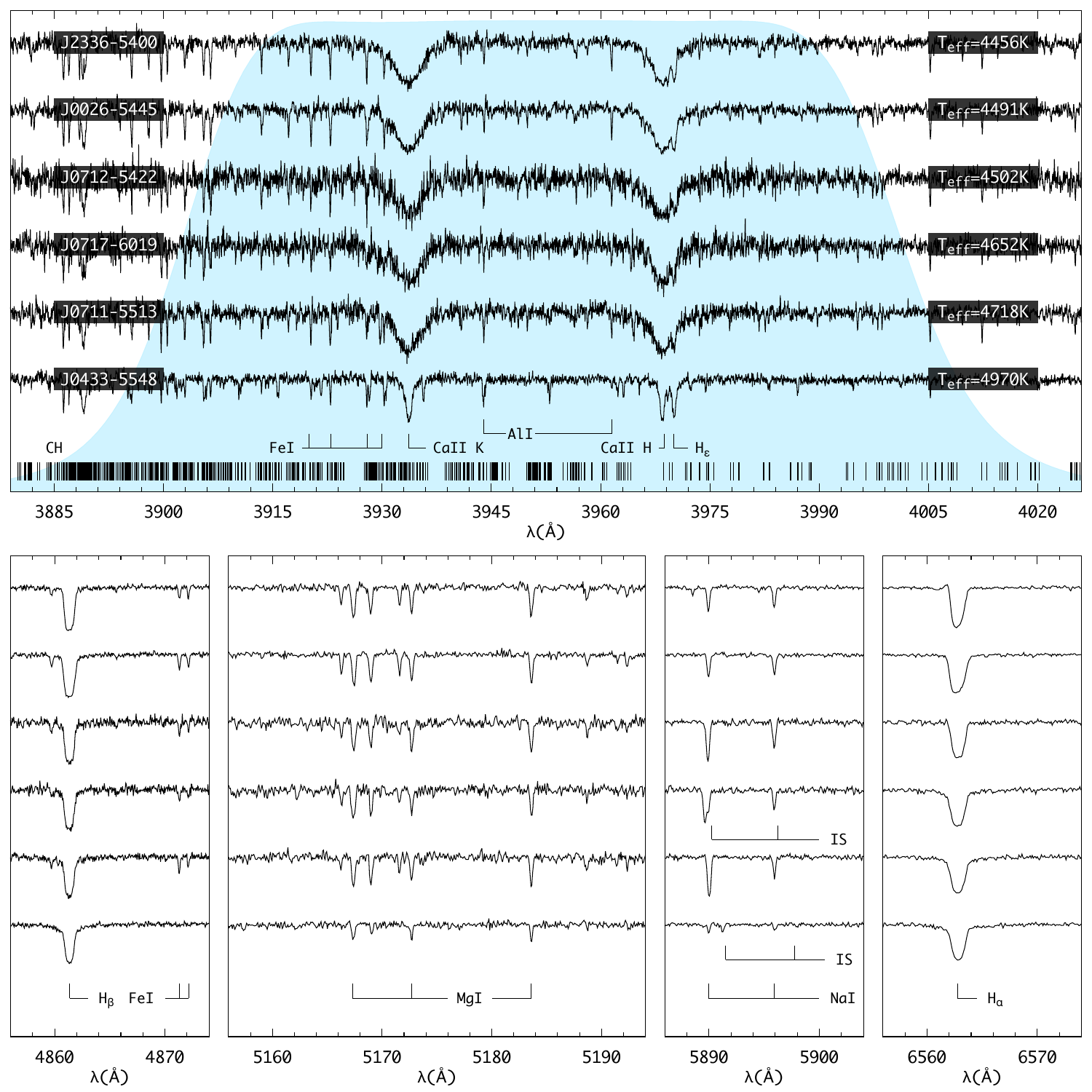}
     \caption{Selected regions of the MIKE spectra for the program stars, sorted by effective temperature. Absorption features of interest are labeled. The shaded area in blue on the top panel is the scaled transmission curve of the DECam N395 filter.}
 \label{spectra}
\end{figure*}

High-resolution spectroscopy for the program stars was obtained on December 30th, 2024 using the Magellan Inamori Kyocera Echelle \citep[MIKE;][]{mike} spectrograph, mounted on the 6.5 m Magellan–Clay Telescope at Las Campanas Observatory. The observing setup included a 1\farcs0 slit with $2\times2$ on-chip binning, yielding a resolving power of $R\sim 28,000$ in the blue spectrum and $R\sim 22,000$ in the red spectrum. Reduction and analysis were performed on a per-exposure basis throughout the night, to assess signal-to-noise ratios (S/N) and exclude any interlopers in our observations. One source (Gaia~DR3~5498603568281416320, \metal$_{\rm MAGIC}=-4.8$), which turned out to be an interloping main-sequence star after analyzing a 16\,min exposure, was excluded from further observations. In summary, six out of the seven stars observed have \metal$\leq-3.0$.
The spectra were reduced using the routines developed for MIKE spectra \citep[CarPy;][]{kelson2003}. Table~\ref{obsinfo} lists IDs, coordinates, magnitudes, and other observational information for the program stars, along with the MAGIC CaHK photometric metallicities. Figure~\ref{spectra} shows selected regions of the high-resolution MIKE spectra (normalized and radial-velocity shifted), including absorption features of interest for stellar parameters (e.g. \ion{Fe}{1}) and chemical abundance (e.g. CH, \ion{Al}{1}, \ion{Mg}{1}, and \ion{Na}{1}) determinations (see Section~\ref{atmparsec} for details). The spectra are sorted, from top to bottom, by increasing effective temperature, which are shown as labels on the right side. From the top panel, it is possible to identify the \ion{Ca}{2}~K line, which is frequently used as a reliable proxy for metallicity estimates in low-resolution spectroscopy and narrow-band photometry. The shaded area represents the transmission curve of the DECam N395 filter.

\begin{figure}[!ht]
 \includegraphics[width=1\linewidth]{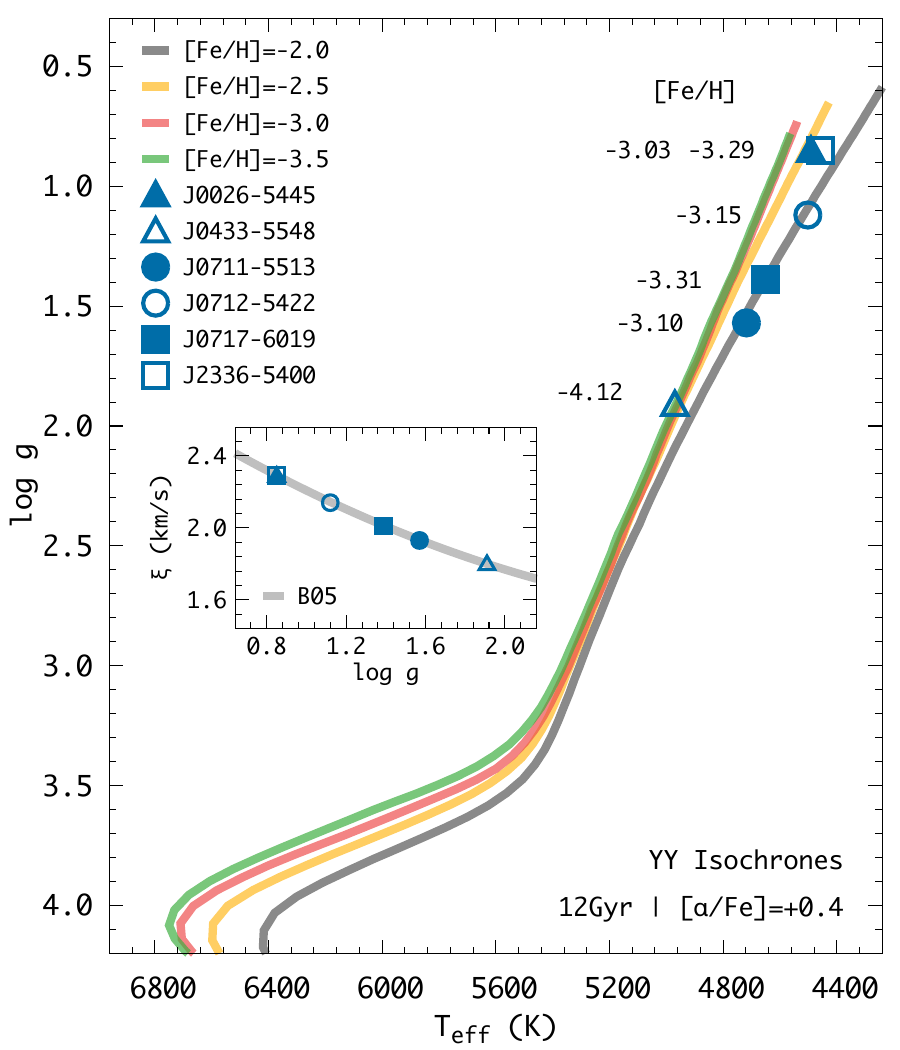}
     \caption{Surface gravity vs. temperature diagram for the program stars, using the parameters listed in Table~\ref{atmpars}. YY Isochrones from \citet{demarque2004} with 12~Gyr, 0.8\Msun, \mbox{[$\alpha$/Fe]=$+0.4$}, and \metal=$-2.0,-2.5,-3.0,-3.5$ are shown for reference. The numbers close to each symbol are the \metal\, values. The inset shows the quadratic relation used for the $\xi$ calculations.}
 \label{isochrone}
\end{figure}

\section{Atmospheric Parameters and Chemical Abundances}
\label{atmparsec}

\subsection{Atmospheric Parameters}

The stellar atmospheric parameters (\teff, \logg, \metal, and $\xi$) for the program stars were determined by a combination of photometric and spectroscopic methods, as described below.
The effective temperatures (\teff) were calculated from the color-\teff-\metal\, relations derived by \citet{mucciarelli21}. The $G$, $BP$, and $RP$ magnitudes were retrieved from the third data release of the Gaia mission \citep[DR3;][]{gaia23dr3} and the $K$ magnitudes from 2MASS \citep{skrutskie2006}. 
The reddening corrections for the Gaia magnitudes were calculated from the iterative procedure outlined in \citet{gaia2018} and the $K$ magnitudes were corrected using the extinction coefficient by \citet{mccall2004}. 
From the magnitudes and their uncertainties, 10$^5$ samples were drawn. For each input color ($BP-RP$, $BP-G$, $G-RP$, $BP-K$, $RP-K$, and $G-K$), the median temperatures were calculated using the giant star relations in \citet{mucciarelli21}. The final \teff\, values for each program star are the weighted mean of the median temperatures for the six different color indices. A metallicity value of \metal=$-3.0$ was used as a first estimate and the temperatures were re-calculated once the spectroscopic \metal\, was determined. The final \teff\, values are listed in Table~\ref{atmpars}.

\begin{deluxetable}{c@{}c@{}c@{}c@{}c@{}r@{}c@{}}[!ht] 
\tabletypesize{\tiny}
\tabletypesize{\footnotesize}
\tablewidth{0pc}
\tablecaption{\label{atmpars} Adopted atmospheric parameters, radial velocities, and S/N for the program stars.}
\tablehead{
\colhead{ID}&
\colhead{\teff}&
\colhead{\logg}&
\colhead{\metal}&
\colhead{$\xi$}&
\colhead{$V_{\rm rad}$}&
\colhead{S/N}\\
\colhead{}&
\colhead{K}&
\colhead{(cgs)}&
\colhead{dex}&
\colhead{\kmsec}&
\colhead{\kmsec}&
\colhead{4500\AA}}
\startdata
\jone   &  4491 & 0.85 & $-$3.03 & 2.29 & 221.7 &  32 \\
\jtwo   &  4970 & 1.91 & $-$4.12 & 1.80 & 269.5 &  29 \\
\jthree &  4718 & 1.57 & $-$3.10 & 1.93 & 317.8 &  18 \\
\jfour  &  4502 & 1.12 & $-$3.15 & 2.14 & 328.6 &  15 \\
\jfive  &  4652 & 1.39 & $-$3.31 & 2.01 & 341.5 &  21 \\
\jsix   &  4456 & 0.85 & $-$3.29 & 2.29 &  70.1 &  30 \\
\enddata
\end{deluxetable}

The surface gravity (\logg) was determined by assuming the program stars are giants and matching their $(g-i)$ color with a 12~Gyr, \metal=$-2.5$, \mbox{[$\alpha$/Fe]=$+0.4$} isochrone from the Dartmouth Stellar Evolution database \citep{dotter2008}. The microturbulent velocity ($\xi$) was calculated using the B05 \citep{barklem2005} quadratic \logg-$\xi$ relation from \citet{ji2023}. The final values are shown in Table~\ref{atmpars}. Figure~\ref{isochrone} shows the stellar parameters for the program stars in comparison with a different set of isochrones (YY - solid lines) from \citet{demarque2004}. These were generated with 12~Gyr, 0.8\Msun, \mbox{[$\alpha$/Fe]=$+0.4$}, and \metal=$-2.0,-2.5,-3.0,-3.5$. The figure inset shows the B05 quadratic \logg-$\xi$ relation from \citet{ji2023}. As expected, there is an excellent agreement between the \logg-\teff\, pairs (from $(g-i)$ colors and Gaia/2MASS magnitudes, respectively) and the YY isochrones.

The metallicity for the program stars (shown as labels on Figure~\ref{isochrone}) was determined spectroscopically from the equivalent widths (EWs) of \ion{Fe}{1} absorption features in the MIKE spectra, by fixing the \teff, \logg, and $\xi$ determined above. Table~\ref{eqwl} in the Appendix lists the lines analyzed in this work, their measured EWs, and the derived chemical abundances. The EWs were obtained by fitting Gaussian profiles to the observed features using the software {\texttt{Spectroscopy Made Harder}} \citep[\texttt{SMHr};][]{casey2014}. The \metal\, values were also calculated using \texttt{SMHr}, which runs the 2017 version of the \texttt{MOOG}\footnote{\href{https://github.com/alexji/moog17scat}{https://github.com/alexji/moog17scat}} code \citep{sneden1973,sobeck2011}, employing one-dimensional plane-parallel model atmospheres with no overshooting \citep{castelli2004}, assuming local thermodynamic equilibrium (LTE). The final \metal\, values, as well as barycentric-corrected radial velocities ($V_{\rm rad}$) and S/N at 4500{\AA}, are provided in Table~\ref{atmpars} for the program stars.

\begin{figure*}
 \includegraphics[width=1\linewidth]{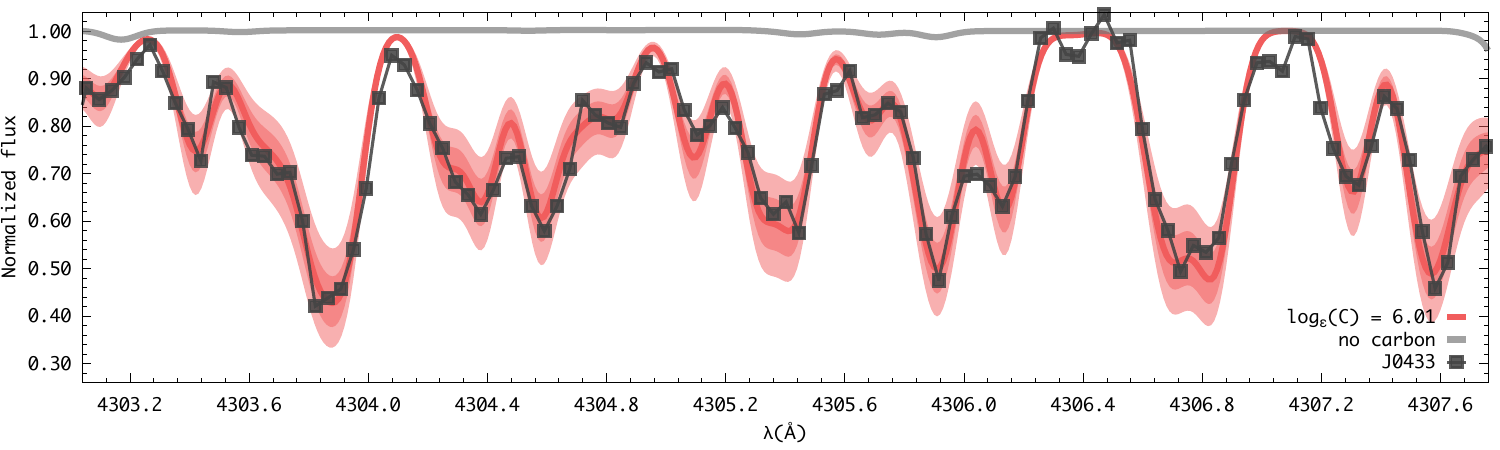}
 \includegraphics[width=1\linewidth]{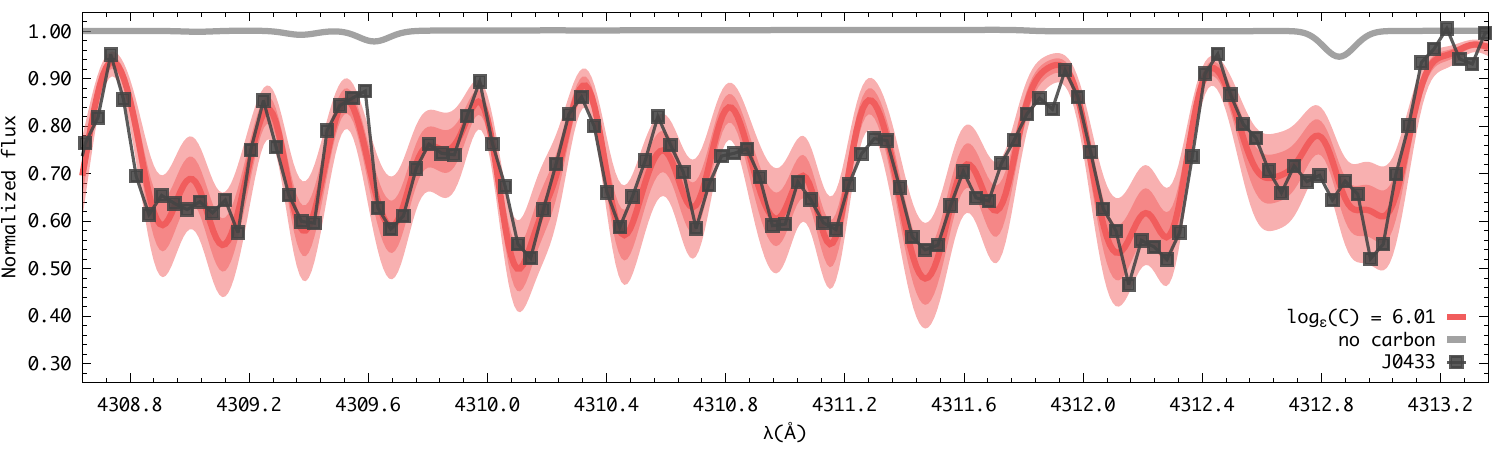}
     \caption{Abundance determination via spectral synthesis for two sections of the Carbon CH G-band for \protect\jtwo. The filled squares connected by the black line represent the MIKE spectrum, the red line is the best fit, and the shaded regions represent $\pm0.1$ and $\pm0.2$~dex from the best-fit abundance. Also shown is a synthetic spectrum without carbon (gray line).}
 \label{csyn}
\end{figure*}

\subsection{Chemical Abundances}

A total of 683 absorption features for 16 elements were identified in the MIKE spectra of the program stars. Abundances were calculated using the 2017 version of \texttt{MOOG} with \texttt{SMHr} from EW and spectral synthesis. EW is mostly used for isolated lines, while synthesis is important for line blends and molecules, or when broadening by isotopic shifts and hyperfine structure splitting has to be accounted for. Table~\ref{eqwl} provides the atomic data, EW values where applicable, and abundances for these features. Abundances determined from spectral synthesis are labeled {\emph{syn}} on the EW column. The line lists used in this work were generated by the \texttt{linemake} code\footnote{\href{https://github.com/vmplacco/linemake}{https://github.com/vmplacco/linemake}} \citep{placco2021}. Logarithmic number abundances ($\log\epsilon$(X)) and abundance ratios (\abund{X}{H} and \xfe{X}) adopt the solar photospheric abundances ($\log\epsilon_{\odot}$\,(X)) from \citet{asplund2009}. Final abundance values and the number of lines measured ($N$) for each element in the program stars are given in Table~\ref{abund}. The $\sigma$ values represent the standard deviation and $\sigma_a$ the adopted uncertainty\footnote{For abundances with $\sigma<0.10$ and $N>1$ in Table~\ref{abund}, we set a standard minimum value of $\sigma_a=0.10$. When $N=1$ for the EW analysis, $\sigma_a=0.15$. For the abundances determined through spectral synthesis, uncertainties were estimated by minimizing the residuals between the MIKE data and a set of synthetic spectra.}.

For the carbon abundance determination, the CH G-band was synthesized in two 4.5\AA-wide regions in the 4300-4315\AA\, range and using a fixed carbon isotopic ratio of $^{12}{\rm C}/^{13}{\rm C}=4$. The best-fit value for each region was determined by minimizing the residuals between the observed and synthetic data. Two examples for the star \jtwo\, are shown in Figure~\ref{csyn}. The filled squares connected by the black line represent the MIKE spectrum, the red line is the best fit, and the shaded regions represent $\pm0.1$ and $\pm0.2$~dex from the best-fit abundance, which is used as the abundance uncertainty. Also shown for reference is a synthetic spectrum without carbon (gray line). Carbon abundances were measured for all six program stars.
Due to their position in the giant branch with \logg~$<2.0$, all the program stars experienced significant carbon depletion in their atmosphere, so evolutionary corrections to the carbon abundances were calculated according to \citet{placco2014c}. The corrections range from $+0.03$ for \jtwo\, to $+0.76$ for \jone.
The average carbon abundances and their corrected values are listed in Table~\ref{abund}. We conservatively set the uncertainties in \cfe\, as 0.20~dex for all program stars.

\begin{deluxetable*}{@{}lrrrrrrrrrrrrrrrrr@{}}[!ht] 
\tabletypesize{\small}
\tabletypesize{\footnotesize}
\tablewidth{0pc}
\tablecaption{Abundances for Individual Species \label{abund}}
\tablehead{
\colhead{Ion}                         & 
\colhead{$\log\epsilon$\,(X)}         & 
\colhead{$\mbox{[X/Fe]}$}             & 
\colhead{$\sigma$}                    & 
\colhead{$\sigma_a$}                  & 
\colhead{$N$}                         &
\colhead{}                            &
\colhead{$\log\epsilon$\,(X)}         & 
\colhead{$\mbox{[X/Fe]}$}             & 
\colhead{$\sigma$}                    & 
\colhead{$\sigma_a$}                  & 
\colhead{$N$}                         &
\colhead{}                            &
\colhead{$\log\epsilon$\,(X)}         & 
\colhead{$\mbox{[X/Fe]}$}             & 
\colhead{$\sigma$}                    & 
\colhead{$\sigma_a$}                  & 
\colhead{$N$}                         }
\startdata 
& \multicolumn{5}{c}{\jone} && \multicolumn{5}{c}{\jtwo} && \multicolumn{5}{c}{\jthree} \\
\hline            
C           &    4.62 & $-$0.78 & \nodata &   0.20 &       1 &&    6.01 & $+$1.70 & \nodata &   0.15 &       1 &&    5.23 & $-$0.10 & \nodata &   0.20 &       1 \\ 
C\vv        &    5.38 & $-$0.02 & \nodata &   0.20 &       1 &&    6.04 & $+$1.73 & \nodata &   0.15 &       1 &&    5.54 & $+$0.21 & \nodata &   0.20 &       1 \\ 
\ion{Na}{1} &    3.09 & $-$0.13 &    0.08 &   0.10 &       2 &&    2.07 & $-$0.05 &    0.09 &   0.10 &       2 &&    3.08 & $-$0.06 & \nodata &   0.15 &       1 \\ 
\ion{Mg}{1} &    4.72 &    0.15 &    0.01 &   0.10 &       3 &&    3.50 &    0.02 &    0.10 &   0.10 &       2 &&    4.66 &    0.16 &    0.06 &   0.10 &       3 \\ 
\ion{Al}{1} &    2.29 & $-$1.13 &    0.18 &   0.13 &       2 &&    1.45 & $-$0.88 & \nodata &   0.15 &       1 &&    2.45 & $-$0.90 &    0.04 &   0.10 &       2 \\ 
\ion{Si}{1} &    4.59 &    0.10 &    0.02 &   0.10 &       2 && \nodata & \nodata & \nodata &\nodata & \nodata &&    4.77 &    0.36 &    0.03 &   0.10 &       2 \\ 
\ion{Ca}{1} &    3.43 &    0.11 &    0.08 &   0.10 &       7 &&    2.28 &    0.06 & \nodata &   0.15 &       1 &&    3.50 &    0.26 &    0.04 &   0.10 &       4 \\ 
\ion{Sc}{2} & $-$0.26 & $-$0.38 &    0.05 &   0.10 &       6 && $-$0.64 &    0.34 & \nodata &   0.15 &       1 &&    0.09 &    0.04 &    0.05 &   0.10 &       5 \\ 
\ion{Ti}{1} &    1.80 & $-$0.12 &    0.03 &   0.10 &       3 &&    1.35 &    0.53 & \nodata &   0.15 &       1 &&    2.11 &    0.26 &    0.07 &   0.10 &       4 \\ 
\ion{Ti}{2} &    1.89 & $-$0.03 &    0.09 &   0.10 &      12 &&    1.23 &    0.40 &    0.05 &   0.10 &       3 &&    2.28 &    0.43 &    0.11 &   0.10 &       9 \\ 
\ion{Cr}{1} &    2.28 & $-$0.33 &    0.05 &   0.10 &       4 &&    1.27 & $-$0.24 & \nodata &   0.15 &       1 &&    2.15 & $-$0.39 &    0.03 &   0.10 &       3 \\ 
\ion{Mn}{1} &    1.27 & $-$1.14 &    0.01 &   0.10 &       3 && \nodata & \nodata & \nodata &\nodata & \nodata &&    1.61 & $-$0.72 &    0.10 &   0.10 &       3 \\ 
\ion{Fe}{1} &    4.47 &    0.00 &    0.19 &   0.19 &     121 &&    3.38 &    0.00 &    0.17 &   0.17 &      34 &&    4.40 &    0.00 &    0.14 &   0.14 &      79 \\ 
\ion{Fe}{2} &    4.48 &    0.00 &    0.06 &   0.10 &       8 && \nodata & \nodata & \nodata &\nodata & \nodata &&    4.54 &    0.13 &    0.02 &   0.10 &       5 \\ 
\ion{Co}{1} &    1.19 & $-$0.77 & \nodata &   0.15 &       1 && \nodata & \nodata & \nodata &\nodata & \nodata &&    1.95 &    0.06 & \nodata &   0.15 &       1 \\ 
\ion{Ni}{1} &    2.93 & $-$0.27 &    0.06 &   0.10 &       2 &&    2.83 &    0.74 & \nodata &   0.15 &       1 &&    3.25 &    0.13 & \nodata &   0.15 &       1 \\ 
\ion{Zn}{1} &    1.38 & $-$0.15 &    0.13 &   0.20 &       2 && \nodata & \nodata & \nodata &\nodata & \nodata &&    2.06 &    0.59 & \nodata &   0.15 &       1 \\ 
\ion{Sr}{2} & $-$1.60 & $-$1.44 &    0.19 &   0.20 &       2 && $-$1.30 & $-$0.05 &    0.03 &   0.20 &       2 && $-$0.21 &    0.02 &    0.02 &   0.20 &       2 \\ 
\ion{Ba}{2} & $-$2.42 & $-$1.57 &    0.03 &   0.20 &       3 && $-$2.67 & $-$0.73 &    0.03 &   0.20 &       2 && $-$1.70 & $-$0.78 &    0.04 &   0.20 &       4 \\ 
\hline
& \multicolumn{5}{c}{\jfour} && \multicolumn{5}{c}{\jfive} && \multicolumn{5}{c}{\jsix} \\
\hline
C           &    4.68 & $-$0.60 & \nodata &   0.20 &       1 &&    5.09 & $-$0.03 & \nodata &   0.20 &       1 &&    4.56 & $-$0.58 & \nodata &   0.20 &       1 \\ 
C\vv        &    5.42 & $+$0.14 & \nodata &   0.20 &       1 &&    5.58 & $+$0.46 & \nodata &   0.20 &       1 &&    5.30 & $+$0.16 & \nodata &   0.20 &       1 \\ 
\ion{Na}{1} &    3.86 &    0.77 & \nodata &   0.15 &       1 &&    3.17 &    0.23 & \nodata &   0.15 &       1 &&    3.09 &    0.14 &    0.07 &   0.10 &       2 \\ 
\ion{Mg}{1} &    4.91 &    0.46 &    0.08 &   0.10 &       4 &&    4.51 &    0.21 &    0.10 &   0.10 &       3 &&    4.78 &    0.47 &    0.14 &   0.14 &       5 \\ 
\ion{Al}{1} &    2.90 & $-$0.40 & \nodata &   0.15 &       2 &&    2.15 & $-$1.00 &    0.03 &   0.10 &       2 &&    2.35 & $-$0.81 &    0.08 &   0.10 &       2 \\ 
\ion{Si}{1} &    4.41 &    0.05 & \nodata &   0.15 &       2 &&    4.67 &    0.47 & \nodata &   0.15 &       1 &&    4.64 &    0.42 &    0.08 &   0.10 &       2 \\ 
\ion{Ca}{1} &    3.60 &    0.41 &    0.10 &   0.10 &       8 &&    3.43 &    0.40 &    0.12 &   0.12 &       5 &&    3.32 &    0.27 &    0.12 &   0.12 &       6 \\ 
\ion{Sc}{2} & $-$0.04 & $-$0.04 & \nodata &   0.15 &       1 && $-$0.12 &    0.04 &    0.10 &   0.10 &       4 && $-$0.27 & $-$0.13 &    0.07 &   0.10 &       3 \\ 
\ion{Ti}{1} &    1.98 &    0.18 &    0.17 &   0.17 &       4 &&    1.81 &    0.16 &    0.08 &   0.10 &       3 &&    1.67 &    0.01 &    0.06 &   0.10 &       4 \\ 
\ion{Ti}{2} &    2.13 &    0.33 &    0.12 &   0.12 &       8 &&    1.84 &    0.19 &    0.10 &   0.10 &       7 &&    2.00 &    0.34 &    0.08 &   0.10 &      15 \\ 
\ion{Cr}{1} &    2.11 & $-$0.38 &    0.13 &   0.13 &       4 &&    2.07 & $-$0.27 & \nodata &   0.15 &       1 &&    2.07 & $-$0.28 &    0.06 &   0.10 &       2 \\ 
\ion{Mn}{1} &    1.18 & $-$1.10 & \nodata &   0.15 &       3 &&    1.19 & $-$0.94 &    0.11 &   0.11 &       3 &&    1.33 & $-$0.81 &    0.06 &   0.10 &       3 \\ 
\ion{Fe}{1} &    4.35 &    0.00 &    0.19 &   0.19 &      73 &&    4.19 &    0.00 &    0.16 &   0.16 &      65 &&    4.21 &    0.00 &    0.17 &   0.17 &      94 \\ 
\ion{Fe}{2} &    4.39 &    0.04 &    0.13 &   0.13 &       5 &&    4.34 &    0.14 &    0.11 &   0.11 &       5 &&    4.25 &    0.04 &    0.08 &   0.10 &       7 \\ 
\ion{Co}{1} &    1.62 & $-$0.22 &    0.14 &   0.14 &       2 &&    1.73 &    0.04 & \nodata &   0.15 &       1 &&    1.57 & $-$0.13 &    0.04 &   0.10 &       3 \\ 
\ion{Ni}{1} &    2.79 & $-$0.28 & \nodata &   0.15 &       1 &&    2.61 & $-$0.30 & \nodata &   0.15 &       1 &&    2.82 & $-$0.11 & \nodata &   0.15 &       1 \\ 
\ion{Zn}{1} &    1.76 &    0.35 & \nodata &   0.15 &       1 && \nodata & \nodata & \nodata &\nodata & \nodata &&    1.51 &    0.24 & \nodata &   0.15 &       1 \\ 
\ion{Sr}{2} & $-$0.46 & $-$0.18 &    0.03 &   0.20 &       2 && $-$0.37 &    0.07 &    0.03 &   0.20 &       2 && $-$0.92 & $-$0.50 &    0.12 &   0.20 &       2 \\ 
\ion{Ba}{2} & $-$1.53 & $-$0.56 &    0.03 &   0.20 &       4 && $-$1.47 & $-$0.34 &    0.06 &   0.20 &       4 && $-$1.54 & $-$0.43 &    0.03 &   0.20 &       4 \\ 
\enddata
\tablenotetext{a}{Using the evolutionary corrections from \citet{placco2014c}.}
\end{deluxetable*}

\begin{figure*}
 \includegraphics[width=0.33\linewidth]{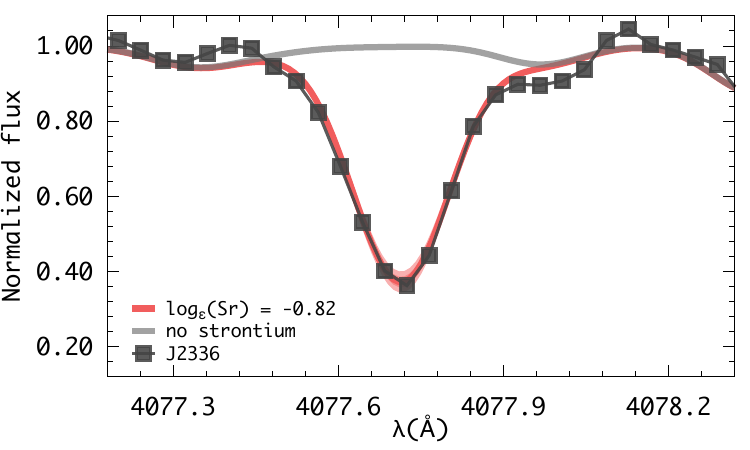}
 \includegraphics[width=0.33\linewidth]{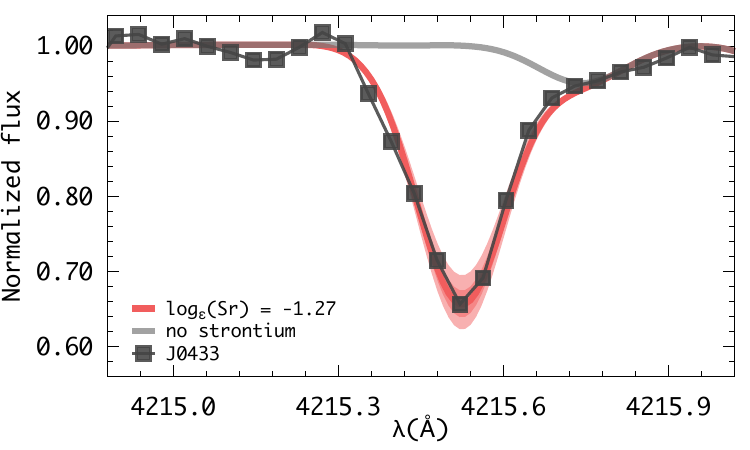}
 \includegraphics[width=0.33\linewidth]{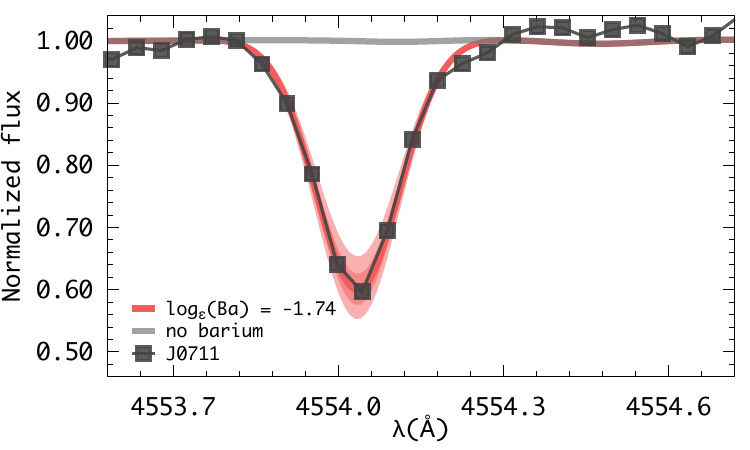}
 \includegraphics[width=0.33\linewidth]{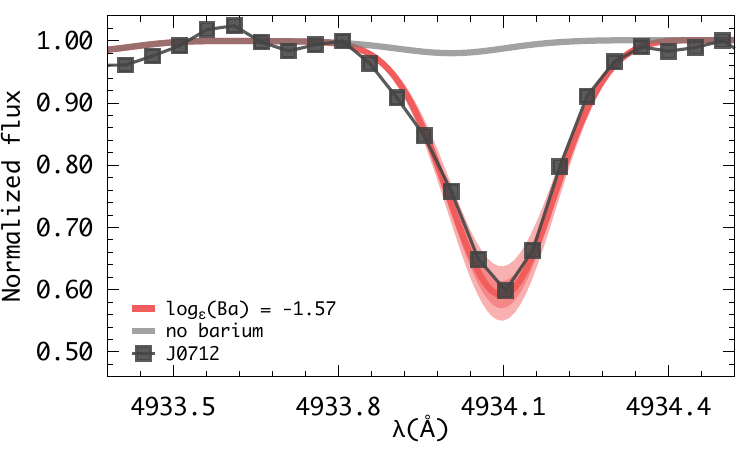}
 \includegraphics[width=0.33\linewidth]{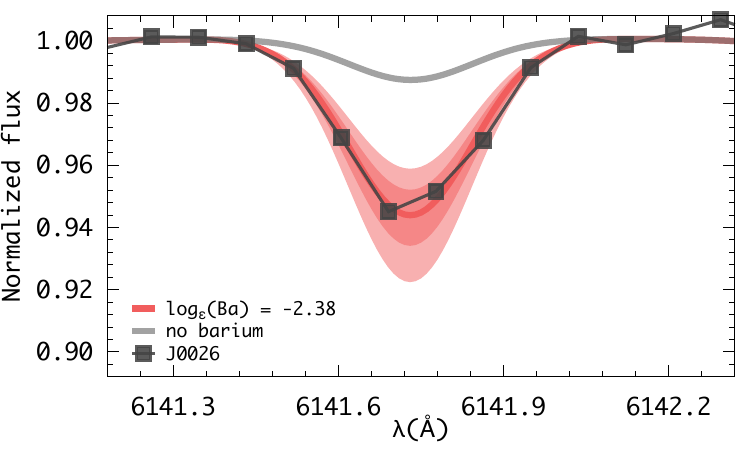}
 \includegraphics[width=0.33\linewidth]{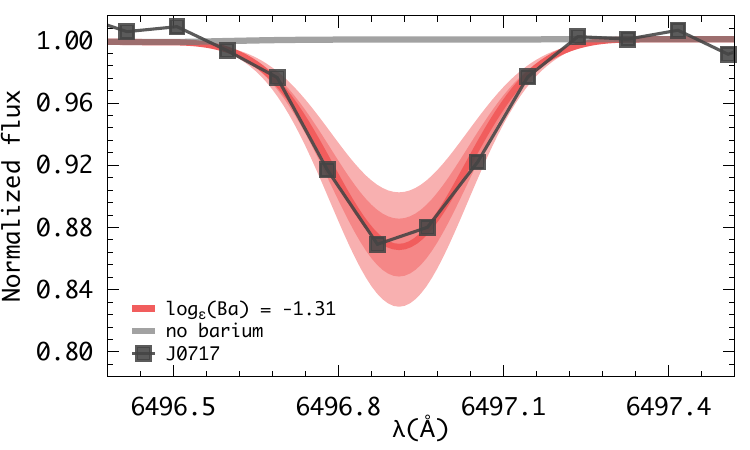}
     \caption{Same as Figure~\ref{csyn}, showing barium and strontium abundance determinations for all six program stars.}
 \label{ncapsyn}
\end{figure*}

For the light elements ($Z<30$) besides carbon, we measured abundances from spectral synthesis for \ion{Al}{1} (3944 and 3961\AA\, - six stars), \ion{Si}{1} (3905 and 4103\AA\, - five stars), \ion{Mn}{1} (4030, 4033, and 4034\AA\, - six stars), and \ion{Zn}{1} (4722 and 4810\AA\, - four stars). The remaining light elements listed in Table~\ref{abund} had their abundances measured through EW analysis. 

Chemical abundances of two heavy elements (strontium and barium) were measured for all six program stars. These include two Sr (4077 and 4215\AA) and four Ba (4554, 4934, 6141, and 6496\AA) absorption features. Figure~\ref{ncapsyn} shows the syntheses for these six absorption features in the program stars. The isotopic fractions for barium were taken from \citet{sneden2008}. For both Sr and Ba, line-by-line abundances for each star agree within 0.15~dex. 

\subsection{Non-LTE Corrections}

Determining Non-LTE (NLTE) corrections is essential to ensuring the accuracy of our abundance determinations and thereby improving the precision of further spectroscopic analysis.
NLTE corrections were obtained for 220 absorption features listed on Table~\ref{abund} using the INSPECT\footnote{\href{http://www.inspect-stars.com/}{http://www.inspect-stars.com/}} database (\ion{Na}{1}), \citet{nordlander2017b} (\ion{Al}{1}), and the MPIA NLTE\footnote{\href{https://nlte.mpia.de/}{https://nlte.mpia.de/}} database (\ion{Mg}{1}, \ion{Si}{1}, \ion{Ca}{1}, \ion{Ti}{1}, \ion{Ti}{2}, \ion{Cr}{1}, \ion{Mn}{1}, \ion{Fe}{1}, \ion{Fe}{2}, and \ion{Co}{1}). The average NLTE corrections for these elements in each of the program stars are given in Table~\ref{nlte}, as well as the number of lines used ($N$). Due to the low temperature of the program stars (\teff$<5000$~K), some elements are heavily affected by NLTE effects. The corrections range from $-0.41$ for \ion{Na}{1} in \jfour\, to $+1.00$ for \ion{Al}{1} in all but one star, with notably high corrections also for \ion{Ti}{1}, \ion{Cr}{1}, and \ion{Mn}{1}.

\begin{deluxetable}{@{}lrrrrr@{}}[!ht]
\tabletypesize{\small}
\tabletypesize{\footnotesize}
\tablewidth{0pc}
\tablecaption{Systematic Abundance Uncertainties for \protect\jone \label{sys}}
\tablehead{
\colhead{Ion}&
\colhead{$\Delta$\teff}&
\colhead{$\Delta$\logg}&
\colhead{$\Delta\xi$}&
\colhead{$\sigma$}&
\colhead{$\sigma_{\rm tot}$\vv}\\
\colhead{}&
\colhead{$+$100\,K}&
\colhead{$+$0.20 dex}&
\colhead{$+$0.15 \kmsec}&
\colhead{}&
\colhead{}}
\startdata
C           &    0.30 & $-$0.08 & $-$0.04 &    0.20 &    0.37 \\
\ion{Na}{1} &    0.10 & $-$0.05 & $-$0.08 &    0.08 &    0.16 \\
\ion{Mg}{1} &    0.10 & $-$0.06 & $-$0.05 &    0.02 &    0.13 \\
\ion{Al}{1} &    0.10 & $-$0.06 & $-$0.07 &    0.18 &    0.23 \\
\ion{Si}{1} &    0.12 & $-$0.04 & $-$0.04 &    0.02 &    0.13 \\
\ion{Ca}{1} &    0.07 & $-$0.03 & $-$0.02 &    0.08 &    0.11 \\
\ion{Sc}{2} &    0.07 &    0.05 & $-$0.04 &    0.05 &    0.11 \\
\ion{Ti}{1} &    0.14 & $-$0.03 &    0.00 &    0.03 &    0.15 \\
\ion{Ti}{2} &    0.05 &    0.05 & $-$0.04 &    0.09 &    0.12 \\
\ion{Cr}{1} &    0.12 & $-$0.04 & $-$0.03 &    0.05 &    0.14 \\
\ion{Mn}{1} &    0.14 & $-$0.06 & $-$0.10 &    0.01 &    0.18 \\
\ion{Fe}{1} &    0.12 & $-$0.04 & $-$0.04 &    0.19 &    0.23 \\
\ion{Fe}{2} &    0.01 &    0.06 & $-$0.03 &    0.06 &    0.09 \\
\ion{Co}{1} &    0.16 & $-$0.03 & $-$0.01 &    0.00 &    0.16 \\
\ion{Ni}{1} &    0.10 & $-$0.03 & $-$0.02 &    0.08 &    0.13 \\
\enddata
\tablenotetext{a}{Calculated from the quadratic sum of the individual error estimates.}
\end{deluxetable}

\subsection{Stellar Parameter Uncertainties}

For this exercise, the abundances determined for the program star \jone\, were used as an example. Systematic uncertainties due to variations in the stellar atmospheric parameters were quantified for the elements with $Z\leq28$ and abundances determined by EW analysis\footnote{We calculated the EW values for the \ion{Al}{1} and \ion{Si}{1} lines for this exercise. Abundances reported in Table~\ref{abund} for these elements were determined from spectral synthesis.} (except for carbon), following \citet{placco2020,placco2023b}.  Table~\ref{sys} shows the variations in chemical abundances when the atmospheric parameters are changed by $+$100~K for \teff\, (typical error from the photometric determinations), $+$0.20~dex for \logg\, (deviation resulting from a 0.1 change in $g-i$ color), and $+$0.15~km\,s$^{-1}$ for $\xi$ (from a 0.2 change in \logg\, in the quadratic relation used for the $\xi$ determination). The $\sigma$ values are the standard error of the mean. For each element, the total uncertainty ($\sigma_{\rm tot}$) is calculated from the quadratic sum of the individual error estimates. 

\section{Analysis and Discussion}
\label{chemod}

\begin{figure*}[!ht]
 \includegraphics[width=1\linewidth]{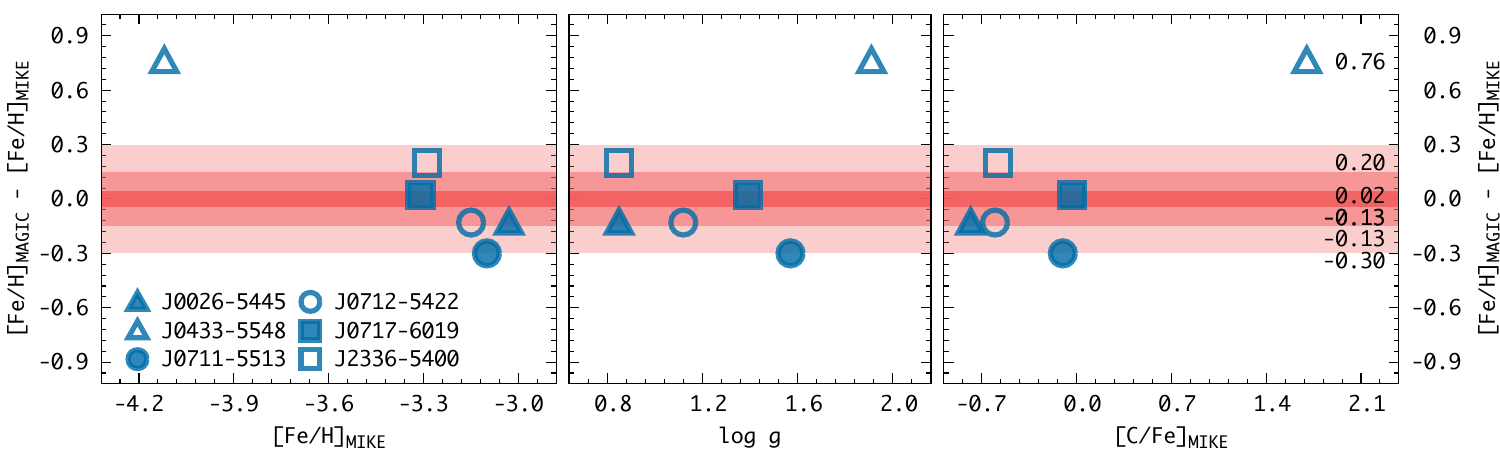}
     \caption{Residual between the photometric and spectroscopic metallicity determinations as a function of \metal$_{\rm MIKE}$ (left), \logg\, (middle), and \cfe\, (right). The shaded areas represent $\pm0.15$ and $\pm0.30$~dex for reference. The symbols are the same as previous figures. The residual values are also shown as labels on the right panel.}
 \label{feh_comp}
\end{figure*}

\subsection{Assessing photometric metallicities in the \metal$\leq-3.0$ regime}
\label{photo}

One of the main advantages of having reliable and well-vetted photometric metallicities is the possibility of bypassing the medium-resolution spectroscopic follow-up and gathering high-resolution data directly. In this section, we assess the accuracy of the MAGIC photometric \metal\, values in the EMP regime. Figure~\ref{feh_comp} shows the behavior of the metallicity residuals ($\Delta$\metal=\metal$_{\rm MAGIC}$$-$\metal$_{\rm MIKE}$) as a function of the high-resolution \metal\, values (left panel), \logg\, (middle panel), and \cfe\, (right panel). The symbols representing the program stars are the same as in previous figures, and the shaded areas show $\pm0.15/\pm0.30$~dex (roughly $1\sigma/2\sigma$) for reference. Also shown in the right panel are the $\Delta$\metal\, values for each star. 

The mean residual value for the five program stars in the $-3.31\leq$\metal$\leq-3.03$ range is $-0.07$~dex, which is within expectations for the EMP regime and does not appear to be correlated with \logg\, or \cfe. For \jtwo, at \metal=$-4.12$, $\Delta$\metal=$+0.76$, meaning the photometric metallicity is overestimated by a significant amount. Even though the \logg\, for this star, is the highest among the program stars (middle panel of Figure~\ref{feh_comp}), its agreement with the isochrones in Figure~\ref{isochrone} when using photometric metallicities suggests it is not the cause for the \metal\, discrepancy. The presence of strong carbon absorption features, however, can lead to a decrease in flux within the narrow-band filter, which translates into a higher N395 magnitude, moving the star downwards in the color-color diagram shown in Figure~\ref{isofeh} and overestimating its \metal. Thus, for \jtwo, its \cfe=$+1.70$ could be the culprit for the metallicity discrepancy.

\begin{figure*}[!ht]
 \includegraphics[width=1\linewidth]{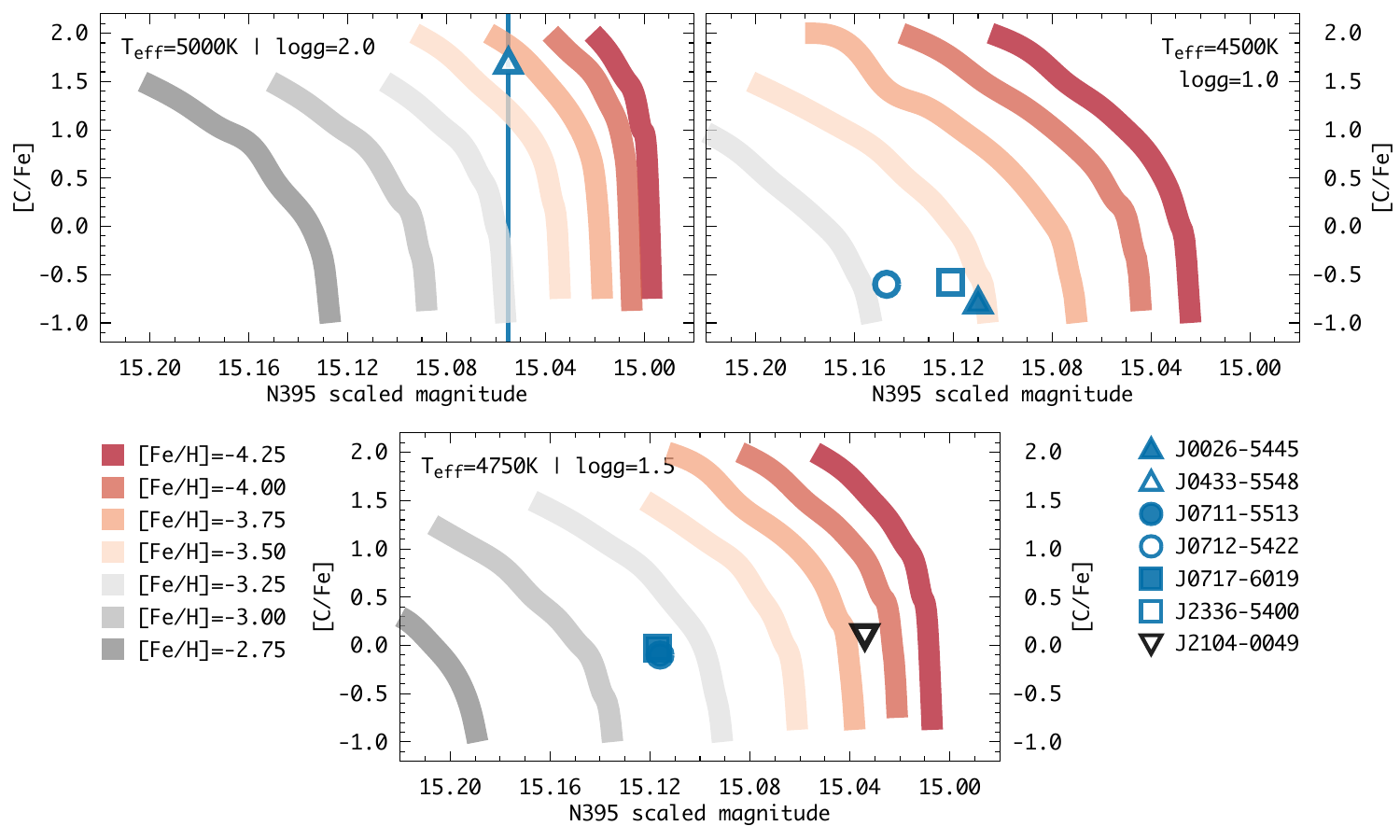}
     \caption{Carbon abundances as a function of scaled N395 magnitudes for different metallicity values (solid lines) for stars with \teff=5000~K and \logg=2.0 (top left panel), \teff=4500~K and \logg=1.0 star (top right panel), and \teff=4750~K and \logg=1.5 (lower panel). The program stars are shown with the same symbols as in previous figures. SPLUS~J2104$-$0049 \citep{placco2021b} is also shown. See the text for further details.}
 \label{mag_cemp}
\end{figure*}

\begin{figure*}[!ht]
 \includegraphics[width=1\linewidth]{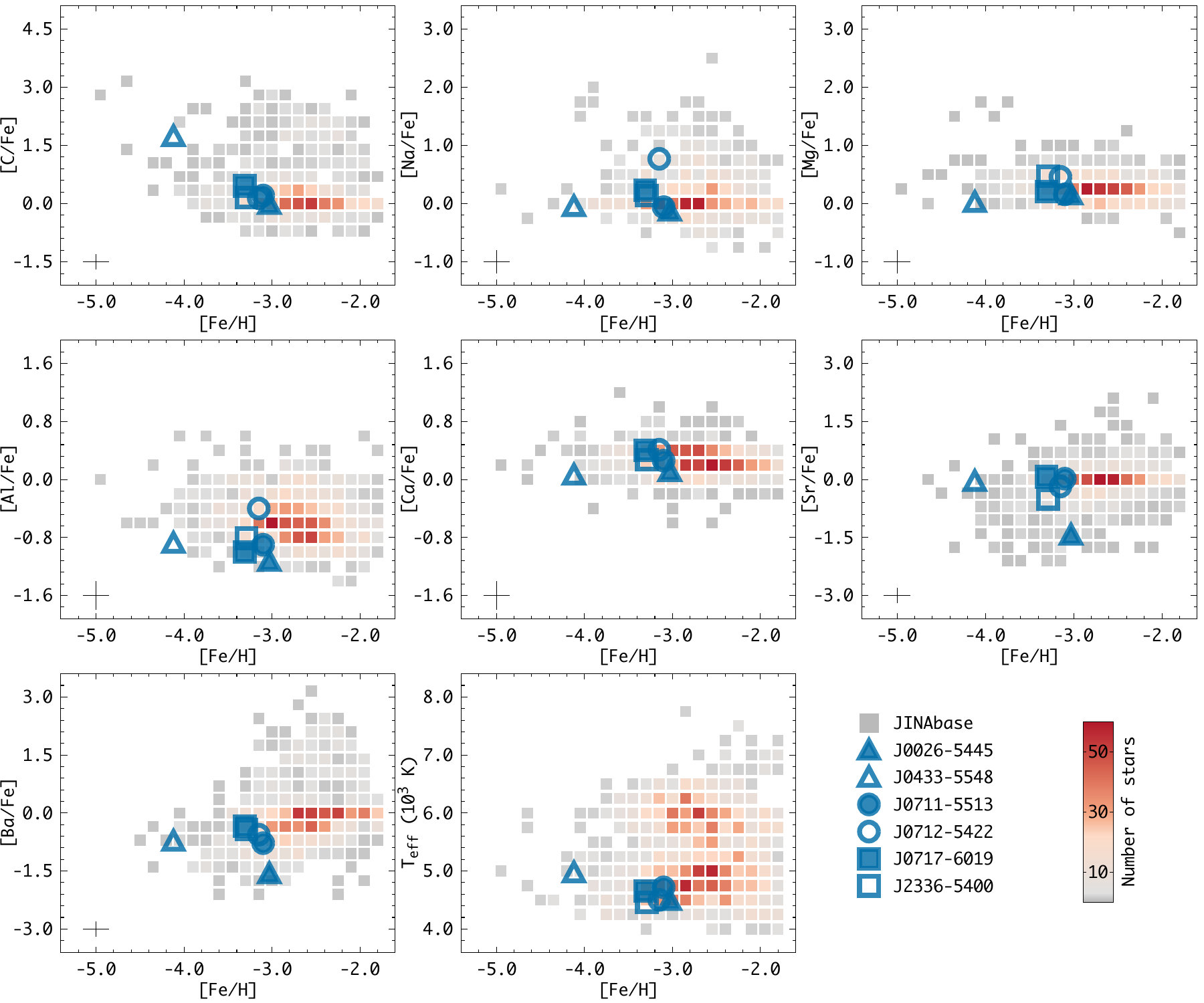}
     \caption{Selected LTE [X/Fe] abundance ratios and \teff\, as a function of the metallicity for the program stars (symbols), compared with the \texttt{JINAbase} \citep{jinabase} literature compilation (density map).}
 \label{jinabase}
\end{figure*}

CH absorption features can be present in the wavelength range probed by the N395 filter for carbon-enhanced, metal-poor giant stars. Because of that, we attempted to quantify the effect of increasing carbon abundances in the N395 magnitude for three cases with varying \metal: \teff(K)/\logg = 5000/2.0, 4750/1.5, and 4500/1.0. These parameters are close to the ones for the program stars, and the results are shown in Figure~\ref{mag_cemp}. The magnitudes were calculated by convolving the normalized flux for a grid of synthetic spectra (with varying \teff, \logg, \metal, and \cfe\, as shown in the Figure) with the N395 narrow band filter transmission curve\footnote{Available at \href{https://noirlab.edu/science/filters/ctdecamn395}{https://noirlab.edu/science/filters/ctdecamn395}}. Each colored line (with 0.01~mag thickness) represents a fixed metallicity and varying carbon abundance. Also shown are the program stars and SPLUS~J2104$-$0049 (\teff=4812~K, \logg=1.95, \metal=$-4.19$, \cfe=$+0.10$) from \citet{placco2021b} as a comparison.

Without a priori knowledge of the carbon abundance, it is possible to see that a star with stellar parameters and N395 magnitude similar to \jtwo\, (\metal=$-4.12$ and \cfe=$+1.70$) could either have \metal=$-3.75$/\cfe=$+1.75$ or \metal=$-3.25$/\cfe$<0.0$ (vertical solid blue line on the top left panel of Figure~\ref{mag_cemp}). The effect is more pronounced for the \teff(K)/\logg=4500/1.0 case; a 2~dex difference in \cfe\, can lead to a 0.5~dex difference in \metal. 
In contrast, SPLUS~J2104$-$0049 has a much lower carbon abundance, and its N395 scaled magnitude better aligns with the \metal$\sim-4.0$ line in the bottom panel. The same applies to the other five program stars (with \cfe$\leq+0.10$), which are all between the \metal$=-3.0$ and $-3.5$ tracks.

From this exercise, we conclude that the enhancement in carbon mimics a higher metallicity and can lead to the discrepancy seen in Figure~\ref{feh_comp}. This can impact CEMP frequencies as a function of \metal\, for samples drawn from photometric metallicities. Additional observations in this metallicity regime will help further assess those differences.

\begin{figure*}[!ht]
 \includegraphics[width=1\linewidth]{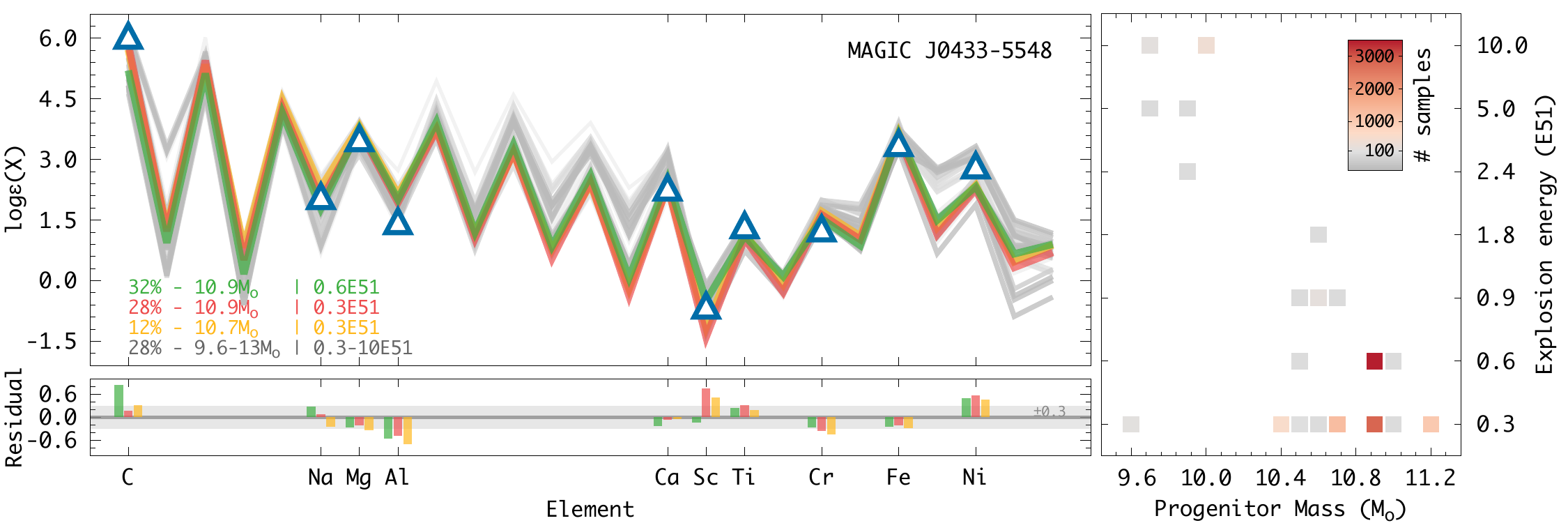}
     \caption{Upper left panel: light-element chemical abundance pattern of \protect{\jtwo}, (open triangles) compared with yields from metal-free supernova models (solid lines). The progenitor mass and explosion energy of the models and their percentage occurrence among the abundance pattern resamples of \protect{\jtwo} are shown as solid lines. Lower left panel: residuals between observations and the three best-fit models. A $\pm$0.3~dex shaded area is shown for reference. Right panel: Density map of progenitor mass and explosion energy combinations for the 10,000 abundance pattern re-samples.}
 \label{starfit}
\end{figure*}

\subsection{Comparison with Chemical Abundance Data from the Literature}
\label{literature}

We compared selected chemical abundances determined for the program stars with data from the latest version\footnote{\href{https://github.com/Mohammad-Mardini/JINAbase-updated}{https://github.com/Mohammad-Mardini/JINAbase-updated}} of the \texttt{JINAbase} literature compilation \citep{jinabase}. This version includes data for the Milky Way halo, as well as classical and
ultra-faint dwarf galaxies. Upper limits in the literature data were excluded, and abundance ratios use the solar atmospheric values from \citet{asplund2009}. The comparison is restricted to the metallicity range $-5.0\leq$\metal$\leq-2.8$. For stars with multiple measurements, the {\texttt{priority==1}} flag was adopted, which selects the most reliable abundances for a given star \citep[see][for further details]{jinabase}. 
Figure~\ref{jinabase} shows this comparison for seven elements (C, Na, Mg, Al, Ca, Sr, and Ba) and \teff. The program stars are presented as symbols, and the \texttt{JINAbase} sample is represented as a density map with the color scale shown in the bottom right part of the figure. 

The program stars appear to follow the general trends for the literature sample, except for Al, where the abundances for the program stars seem lower. This could be due to the lower temperature of the program stars when compared with the literature compilation (center-bottom panel of Figure~\ref{jinabase}), which makes NLTE corrections for Al larger. \jfour\, shows Na and Mg abundances that are 0.2-0.5~dex higher than the average for our sample. The largest difference within the program stars is for carbon (top left panel), which can be explained by the arguments presented in Section~\ref{photo}. The high \cfe\, value for \jtwo\, agrees with the general trends for stars in the \metal$\leq-4.0$ regime. For the heavy elements Sr and Ba, \jone\, shows abundance ratios roughly 1~dex lower than the other program stars but still within the literature values. Its metallicity (\metal=$-3.03$) and neutron-capture element abundance ratio (\abund{Sr}{Ba}=$+0.13$) are consistent with values for metal-poor stars in ultra-faint dwarf galaxies \citep{casey2017,roederer2017} and can help distinguish between formation processes that contribute to the enrichment in ultra faint dwarf galaxy analogs. We provide further insights on the possible origin of \jone\, in Section~\ref{kinsec}.

\subsection{\protect\jtwo: Insights on the progenitor population of UMP stars}
\label{jtwo}

 Due to its low metallicity and high carbon abundance, \jtwo\, can be considered a bonafide second-generation star, potentially formed from a gas cloud polluted by a single Pop. III supernova explosion in the early universe. With \xfe{Ba}=$-0.73$, \jtwo\, is classified as a Group~III CEMP-no \citep{frebel2015, yoon2016} star. It also shows a remarkably low magnesium-to-carbon ratio, \abund{Mg}{C}=$-$1.71, which places it in the ``mono-enriched'' regime proposed by \citet{hartwig2018}.

With the chemical abundances determined in this work, we repeated the exercise outlined in \citet{placco2024} by using the \texttt{starfit}\footnote{\href{https://starfit.org/}{https://starfit.org/}} code, containing Pop. III supernova nucleosynthesis yields from \citet{heger2010}. This particular type of progenitor has a pristine Big Bang nucleosynthesis composition, and the models ignore the effects of mass loss and rotationally induced mixing during the star's evolution. We have used the \texttt{znuc2012.S4} nucleosynthesis yields for 16,800 models with progenitor masses from 9.6~\Msun\, to 100~\Msun, explosion energies in the $0.3-10\times10^{51}$~erg range, and a mixing parameter (0.0 to 0.25). We refer the reader to \citep{heger2002,heger2010} for further details.

We re-sampled the chemical abundances for the ten elements with $Z<30$ measured in \jtwo\, (Table~\ref{abund}) 10,000 times, assuming Gaussian distributions centered in the \eps{X} values and with a fixed standard deviation $\sigma=0.2$, to account for the systematic uncertainties presented in Table~\ref{sys}. Figure~\ref{starfit} shows the results of this exercise. The symbols in the top-left panel are the chemical abundances for \jtwo\, and the solid colored lines represent the best-fit models, with their percentage occurrence, progenitor mass, and explosion energy provided as labels. The bottom-left panel shows the residuals (observation minus model abundance) for the three most frequent best-fit models. The right panel shows the density map (progenitor mass vs. explosion energy) of the 10,000 best-fit models. It is possible to see that all best-fit models are within a narrow range of progenitor masses ($9.6\leq$\Msun$\leq11.2$), with explosion energies of $E\leq0.6\times10^{51}$~erg in roughly 95\% of the re-samples.

\begin{figure}[!ht]
 \includegraphics[width=1\linewidth]{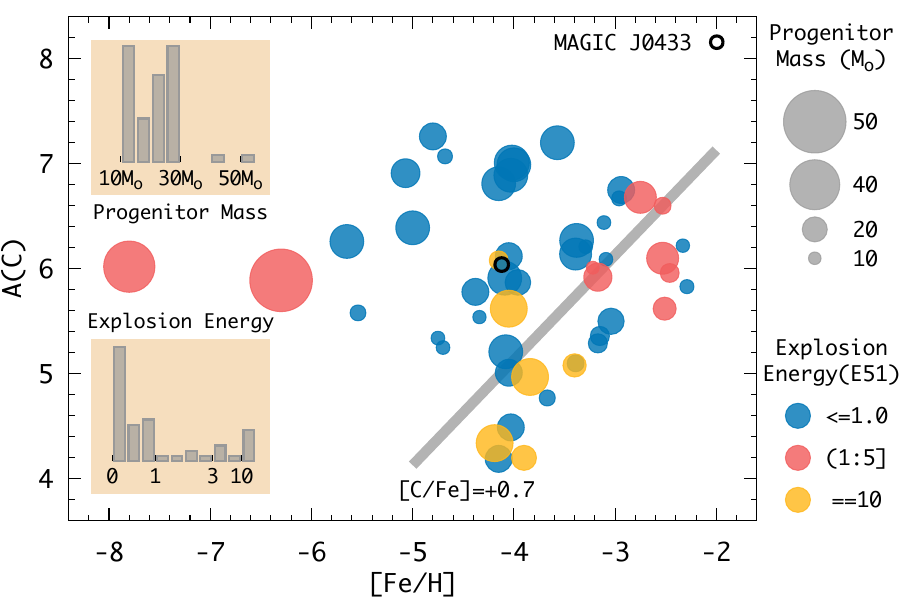}
     \caption{Absolute carbon abundances as a function of the metallicity for stars in the literature where \texttt{starfit} results are available, including \protect\jtwo\, (blue filled circle with a black outline). The point sizes and colors reflect the progenitor mass and explosion energies, and the insets show the progenitor mass (top left) and explosion energy (bottom left) distributions for the 53 stars. Individual references include \citet{andales2024,dasilva2025,dovgal2024,ezzeddine2019,frebel2015b,frebel2019,hansen2020,hansen2014,jeong2023,lagae2023,mardini2019b,mardini2020,mardini2022,mardini2024,placco2014b,placco2015,placco2016,placco2016b,placco2020,placco2021b,placco2023b,placco2024,roederer2016,skuladottir2021,skuladottir2024,webber2023,zaremba2025}.}
 \label{acfeh}
\end{figure}

To place the results above in context, we searched the literature for studies using the \texttt{starfit} code to assess the progenitor population of low-metallicity stars. Even though we acknowledge that the models in \texttt{starfit} provide only one possible progenitor type for Pop.~III, comparing a group of stars analyzed using the same methodology can be insightful. We found 52 stars in the literature with available metallicities (\metal$\leq-2.0$), $A(C)$\footnote{Absolute carbon abundance; $A(C)=\log(N_C/N_H)+12$.}, progenitor masses, and explosion energies. Figure~\ref{acfeh} shows the $A(C)$ values as a function of \metal\, for the literature sample, including \jtwo\, (blue filled circle with a black outline). Upper limits are included in the figure. The point size is proportional to the progenitor mass, the point colors denote the range in explosion energy, and the solid line is the CEMP star threshold defined by \citet{aoki2007}. The insets show the distributions of progenitor mass (top left) and explosion energy (bottom left) for the 53 stars. Individual references are shown in the figure caption. 

The first notable aspect in Figure~\ref{acfeh} is that the two lowest metallicity stars in the sample, SMSS~J0313$-$6708 at \metal$<-7.8$ \citep{bessell2015} and SDSS~J0023$+$0307 at \metal$<-6.3$ \citep{frebel2019}, have the highest progenitor masses at 40\Msun\, and 50\Msun, respectively. Apart from these, all other stars have progenitor masses in the 10-30\Msun\, range. Also worth pointing out is that, in the $-6\leq$\metal$\leq-2$ range, 85\% (28 out of 33 stars) of the progenitors for the CEMP stars (points above the solid line at \cfe$=+0.7$) have explosion energies $\leq1.0$E51. For the non-CEMP stars, the fraction decreases to 55\% (10 out of 18 stars), which corroborates the hypothesis that higher explosion energies would increase the availability of iron (and other metals) for a subsequent star formation episode. Finally, \jtwo\, shows similar progenitor mass and explosion energy as other stars with similar metallicities.

\citet{ishigaki2018} compared the chemical abundance patterns of about 200 EMP stars with metal-free mixing-fallback models for Pop.~III stars and found that progenitor masses are predominantly $M\leq40$\Msun, with half of their sample having their best fit from a high explosion energy (10E51), 25\Msun\, hypernova. Future observations of EMP and UMP stars will help further constrain the mass distribution and initial mass function of the first stars \citep{fraser2017,hartwig2024} and develop a better understanding of the environments that harbored such extreme objects.

\begin{figure*}[!ht]
 \includegraphics[width=1\linewidth]{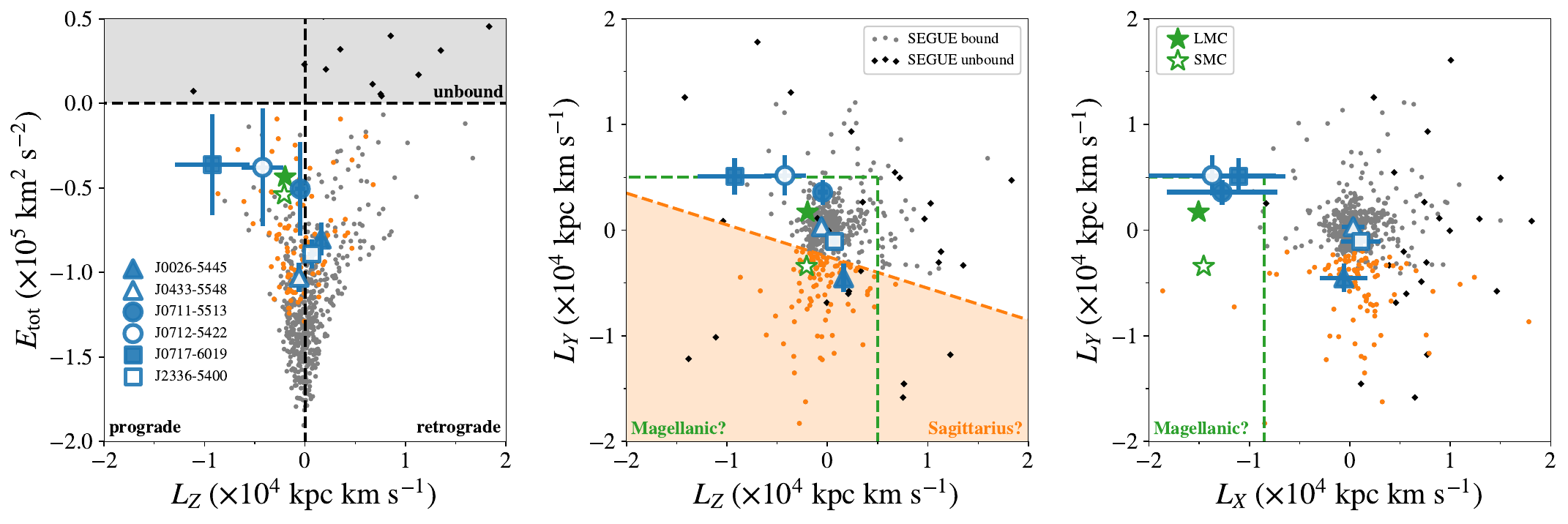}
     \caption{Left: $(L_Z,E_{\rm tot})$. Stars with $L_Z < 0$ are on prograde motion while those exhibiting $L_Z > 0$ are retrograde. The gray band at $E_{\rm tot} > 0$ highlights the threshold for unbound stars under the \citet{mcmillan2017} Milky Way model potential. Middle: $(L_Z,L_Y)$. The orange polygon highlights the region of parameter space occupied by stars likely associated with the Sagittarius dSph and stream \citep{johnson2020sgr}. Those stars from the SEGUE low-metallicity catalog (see text) that occupy this kinematic locus are shown as orange dots. Right: $(L_X,L_Y)$. The bottom left corner of both middle and right panels (green dashed lines) might plausibly host Magellanic stellar debris \citep{chandra2023}. In all panels, as in previous figures, our program stars from MAGIC are plotted as blue and/or white symbols. Gray dots and black diamonds are the rest of the stars from the SEGUE low-metallicity sample that are either bound or unbound, respectively.}
\label{kindyn}
\end{figure*}

\begin{deluxetable*}{@{}rrrrrrrrc@{}}[!ht] 
\tabletypesize{\tiny}
\tabletypesize{\footnotesize}
\tablewidth{0pc}
\tablecaption{\label{kintable} Kinematic/dynamical information for the program stars in the \citet{mcmillan2017} model.}
\tablehead{
\colhead{ID}&
\colhead{$L_X$}&
\colhead{$L_Y$}&
\colhead{$L_Z$}&
\colhead{$r_{\rm peri}$}&
\colhead{$r_{\rm apo}$}&
\colhead{$e$}&
\colhead{$E_{\rm tot}$}&
\colhead{$f_{\rm LMC}$}\\
\colhead{}&
\colhead{(${\rm kpc}\,{\rm km}\,{\rm s^{-1}}$)}&
\colhead{(${\rm kpc}\,{\rm km}\,{\rm s^{-1}}$)}&
\colhead{(${\rm kpc}\,{\rm km}\,{\rm s^{-1}}$)}&
\colhead{(kpc)}&
\colhead{(kpc)}&
\colhead{}&
\colhead{(${\rm km}^2\,\,{\rm s}^{-2}$)}&
\colhead{}
}
\startdata
\jone   & $-$599$^{+2204}_{-2492}$    & $-$4519$^{+1348}_{-1390}$ & 1613$^{+211}_{-198}$      & 13$^{+7}_{-2}$   & 63$^{+13}_{-15}$    & 0.65$^{+0.05}_{-0.06}$ & $-$80,266$^{+8568}_{-10{,}639}$ & $\dots$ \\
\jtwo   & 338$^{+760}_{-1438}$        & 330$^{+1007}_{-547}$    & $-$587$^{+342}_{-534}$    & 2$^{+2}_{-1}$    & 40$^{+8}_{-7}$      & 0.89$^{+0.04}_{-0.08}$ & $-$102,884$^{+8666}_{-8578}$ & $\dots$ \\
\jthree & $-$12,715$^{+4541}_{-6503}$ & 3590$^{+1224}_{-1120}$  & $-$476$^{+467}_{-528}$    & 40$^{+9}_{-21}$  & 110$^{+182}_{-71}$  & 0.49$^{+0.25}_{-0.16}$ & $-$50,647$^{+29{,}315}_{-26{,}269}$ & 75.4\% \\
\jfour  & $-$13,705$^{+5615}_{-7037}$ & 5173$^{+2027}_{-1780}$  & $-$4226$^{+1830}_{-2301}$ & 38$^{+11}_{-33}$ & 128$^{+418}_{-101}$ & 0.56$^{+0.28}_{-0.18}$ & $-$37,847$^{+36{,}503}_{-33{,}131}$ & 48.6\% \\
\jfive  & $-$11,048$^{+3972}_{-5258}$ & 5098$^{+1876}_{-1599}$  & $-$9229$^{+3407}_{-3963}$ & 37$^{+9}_{-28}$  & 156$^{+312}_{-117}$ & 0.61$^{+0.21}_{-0.13}$ & $-$36,219$^{+32{,}658}_{-27{,}301}$ & 20.1\% \\
\jsix   & $-$1090$^{+1492}_{-2452}$   & $-$1055$^{+450}_{-750}$ & 699$^{+282}_{-635}$       & 5$^{+3}_{-2}$    & 54$^{+11}_{-12}$ & 0.84$^{+0.07}_{-0.13}$ & $-$89,204$^{+7911}_{-9205}$ & $\dots$ \\ \hline
LMC & $-$15,061$^{+439}_{-455}$   & 1711$^{+267}_{-304}$    & $-$1974$^{+431}_{-427}$ & 47$^{+1}_{-1}$ & 172$^{+23}_{-16}$ & 0.57$^{+0.04}_{-0.03}$ & $-$43,519$^{+2782}_{-2533}$ & $\dots$ \\
SMC & $-$14,545$^{+1359}_{-1596}$ & $-$3431$^{+392}_{-378}$ & $-$2068$^{+403}_{-551}$ & 60$^{+2}_{-2}$ & 100$^{+30}_{-20}$ & 0.25$^{+0.10}_{-0.09}$ & $-$54,045$^{+6085}_{-5187}$ & $\dots$ \\
\enddata
\end{deluxetable*}

\begin{figure*}[!ht]
 \includegraphics[width=0.3333\linewidth]{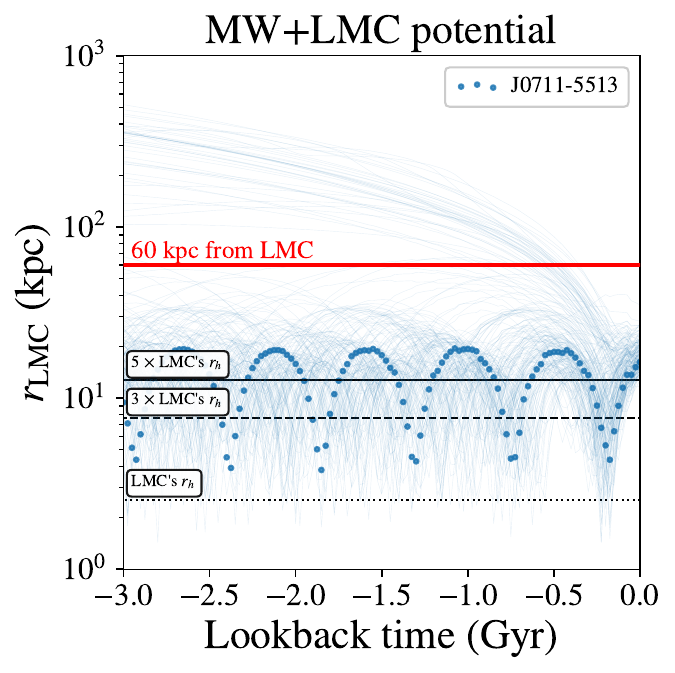}
 \includegraphics[width=0.3333\linewidth]{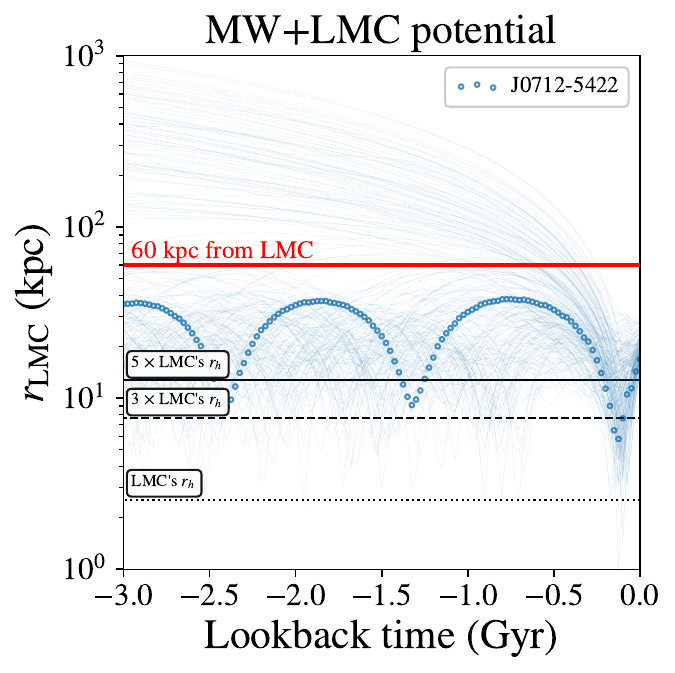}
 \includegraphics[width=0.3333\linewidth]{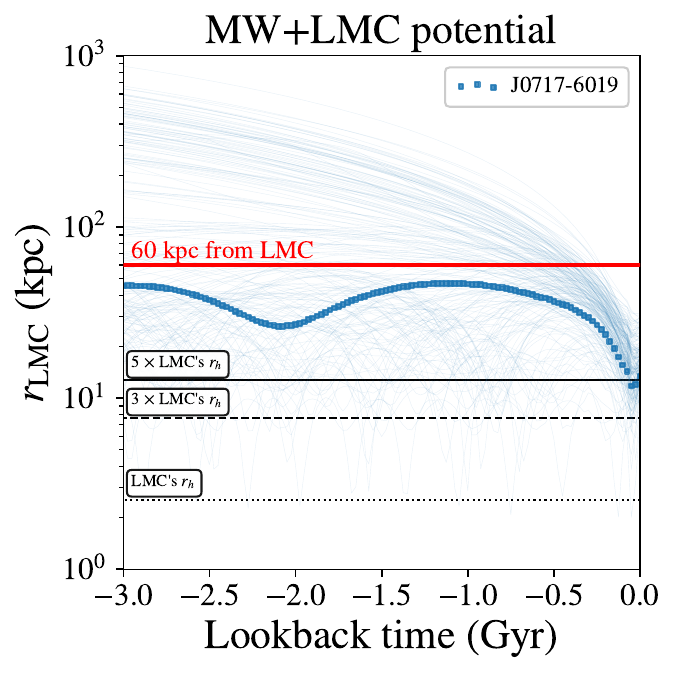}
     \caption{\gui{Orbits of \jthree \ (left), \jfour \ (middle), and \jfive \ (right) in a model potential that includes both the Milky Way and the LMC on first infall \citep{vasiliev2021sgr}. The panels show the different star's distance to the LMC ($r_{\rm LMC}$) as a function of lookback time. The thin blue lines depict 200 realizations for each of these stars' orbit (see text). Dotted, dashed, and solid black horizontal lines represent 1$\times$, 3$\times$, and 5$\times$ the LMC's circularized half-light radius ($r_h \approx 2.5\,{\rm kpc}$; \citealt{pace2024lvdb}).}
     }
\label{kindyn_lmc}
\end{figure*}

\subsection{Kinematics and dynamics}
\label{kinsec}

On the chemical abundance front, with the exception of low Sr and Ba for \jone, our sample of EMP/UMP stars broadly follows the trends of halo stars compiled in \texttt{JINAbase}. 
Beyond their intrinsic chemical properties, the orbital characteristics of our target stars provide crucial insights into their dynamical histories and possible associations with known stellar streams and disrupted satellite systems, reinforcing their significance in the broader narrative of Galactic structure and evolution.
Therefore, in this section, we look at the kinematics and dynamics of our program stars to gain further insights into their origins. On-sky positions and proper motions are taken from the Gaia DR3 astrometric catalog. Line-of-sight radial velocities ($V_{rad}$ in Table \ref{atmpars}) are derived from our MIKE spectra (Section \ref{observations}). The typical $V_{\rm rad}$ uncertainty is $\sim$2\,\kmsec \ at the S/N achieved by our MIKE spectra (Table \ref{atmpars}). Distance moduli in the MAGIC catalog\footnote{MAGIC internal collaboration catalog version \texttt{v250130}.} are obtained iteratively by matching observed DECam $g$ magnitudes and ($g-i$) colors with metal-poor Dartmouth isochrones after an initial guess based on Gaia DR3's parallaxes. All of our sample stars are located at heliocentric distances $35 < d_{\rm helio}/{\rm kpc} \lesssim 55$, hence probing the most ancient portion of the stellar population residing in the outer regions of the Milky Way's halo.

With the above-described 6D phase-space information at hand, we compute kinematic and dynamical quantities of interest. We verified that between the most metal-poor isochrones (${\rm [Fe/H]} = -2.5$, $-3.0$, and $-3.5$) available in the MESA Isochrones and Stellar Tracks (MIST) set of models \citep{Dotter+2016, Choi+2016}, there is a $\sim$0.4\,mag difference in distance modulus within the DECam $g-r$ color interval occupied by our sample stars. This translates into a $\sim$20\% error in heliocentric distances, which we assume for our calculations; this exact choice does not affect our interpretations. As a sanity check, we have also verified the most metal-poor models available in the Bag of Stellar Tracks and Isochrones \citep[BaSTI;][]{hidalgo2018basti, Pietrinferni2021basti}. The difference between $\rm[Fe/H]=-2.5$ and $-3.2$ (12.5\,Gyr, $\rm[\alpha/Fe] = +0.4$) models is less extreme in BaSTI, only $\sim$0.1\,mag in the RGB, which translates to just $\sim$5\% distance difference at 50\,kpc. Therefore, we are confident that our adopted fractional uncertainty for distances is conservative.  

The adopted position and velocity vectors of the Sun in the Cartesian Galactocentric frame are $(X,Y,Z)_\odot = (-8.2, 0.0, 0.0)$\,kpc, following the recommendation by \citet{hawthorn2016}, and $(V_X,V_Y,V_Z)_\odot = (11.10, 245.04, 7.25)$\,km\,s$^{-1}$, with the peculiar motion from \citet{Schonrich+2010} and the local standard of rest from \citet{mcmillan2017}, respectively. Specifically for our discussions, we utilize the full angular momentum components $L_X = Y V_Z - Z V_Y$, $L_Y = Z V_X - X V_Z$, and $L_Z = X V_Y - Y V_X$ (Figure \ref{kindyn}); prograde motion is defined by $L_Z < 0$.

We estimate total orbital energies $E_{\rm tot} = \frac{1}{2}(V_X^2 + V_Y^2 + V_Z^2) + \Phi (\mathbf{r})$ for our sample stars adopting the \citet{mcmillan2017} Milky Way axisymmetric model potential $\Phi$ (total mass $M_{\rm MW} = 1.3 \times 10^{12}\,M_\odot$), where $\mathbf{r} = (X,Y,Z)_\star$ is the position vector for a given star $\star$. For each object, we initialized 200 initial conditions sampled from the observed parameters and their (Gaussian) errors. We integrate orbits for 20\,Gyr forward for all stars for each realization using the \texttt{AGAMA} package \citep{vasiliev2019}. We obtain nominal values for kinematic/dynamical quantities, e.g., orbital peri/apocenters as well as $E_{\rm tot}$ itself, as the medians from the resulting distributions with 16$^{\rm th}$/84$^{\rm th}$ percentiles as uncertainties (Table \ref{kintable}). 
For comparison with our program EMP/UMP targets, we perform all the same calculations for a homogeneous sample of low-metallicity stars (\metal$\leq-2.5$) from the Sloan Extension for Galactic Understanding and Exploration \citep[SEGUE;][and \citealt{lee2008a, lee2008b} for stellar parameters]{yanny2009, rockosi2022} survey with available spectro-photometric distances from \texttt{StarHorse} code \citep{starhorse2016original, Queiroz2018starhorse, starhorse2023spec}, which includes $\mathcal{O}(10^2)$ stars at $>$35\,kpc from the Sun. We also compute kinematics for both Large and Small Magellanic Clouds (LMC and SMC) with phase-space information compiled in the Local Volume Database\footnote{\url{https://github.com/apace7/local_volume_database}. References for LMC and SMC data are \citet{vandermarel2002lcm, kallivayalil2013lmc, Pietrzynski2019lmc} and \citet{cioni2000smc, harris2006smc, Zivick2018smc}, respectively.} \citep{pace2024lvdb}.

Figure \ref{kindyn} shows our results for the kinematic/dynamical analyses. A first noteworthy feature of our program targets is that half of them (3/6 stars) exhibit relatively high total orbital energies ($E_{\rm tot} \gtrsim -0.5 \times 10^5\,{\rm km}^2\,\,{\rm s}^{-2}$). These stars are exactly the ones that have the highest $V_{\rm rad}$ of $>$300\,\kmsec \ (Table \ref{atmpars}) in the sample and are located at $45 < d_{\rm helio}/{\rm kpc} \leq 50$. Moreover, these three stars exhibit prograde motion ($L_Z < 0$), while the opposite behavior is seen in the similarly high-energy SEGUE stars, where most of them (28/42) are counter-rotating ($L_Z > 0$). We caution that Milky Way halo samples at $>$35\,kpc are small and the surveys analyzed here, MAGIC and SEGUE, cover completely different regions of the sky, each carrying their own selection biases.

Out of known outer halo substructures, the three high-energy MAGIC stars show $L_Y$ values inconsistent with the Sagittarius stream \citep[][see also \citealt{majewski2000}]{johnson2020sgr}, but plausibly compatible, within $1\sigma$ uncertainties, with the proposed Magellanic stellar debris \citep[][see also \citealt{Medina2023rrl}]{chandra2023}. Their large $L_X \gtrsim 10^4\,{\rm kpc}\,{\rm km}\,{\rm s^{-1}}$ values also support the latter possibility. It might be the case that their kinematics are heavily affected by the perturbation caused by the infall of the Magellanic system \citep[LMC$+$SMC$+$other satellites;][for a review]{vasiliev2023}. Their $L_X$ and $L_Y$ signs also make them incompatible with both Cetus \citep{newberg2009} and Orphan \citep{belokurov2007} streams; see section 2.1 of \citet{naidu2022} and the appendix in \citet{limberg2023}. 

\gui{Given the hint that the high-energy EMP stars found here could be connected to a Magellanic origin, we conduct a Monte Carlo experiment to test this hypothesis. We calculate $10^3$ realizations of these three stars' orbits for 3\,Gyr backward in a time-varying gravitational potential model that includes both the Milky Way (similar implementation to \citealt{mcmillan2017}, but with a triaxial halo) and a first-infall LMC \citep[see][]{besla2007lmc} with total mass $M_{\rm LMC} = 1.5 \times 10^{11}\,M_\odot$ \citep{vasiliev2021sgr}; see Figure \ref{kindyn_lmc}. Whenever an orbit approaches the LMC within $<$60\,kpc at $\geq$2\,Gyr ago, we consider it to be bound (\citealt{chandra2023} criterion). The fraction of realizations for \jthree, \jfour, and \jfive \ that end up bound to the LMC are $f_{\rm LMC} =75.4\%$, 48.6\%, and 20.1\% (see Table \ref{kintable}), respectively, reinforcing a Magellanic association.}

The remaining three lower energy ($E_{\rm tot} < -0.8 \times 10^5\,{\rm km}^2\,\,{\rm s}^{-2}$) stars in our program display more ``standard'' halo-like kinematics. Stars \jsix\, and UMP \jtwo\, have extremely high orbital eccentricities $e = (r_{\rm apo} - r_{\rm peri})/(r_{\rm apo} + r_{\rm peri})$ of $>$0.8 (in \citealt{mcmillan2017} Milky Way-only model potential), which is reminiscent of the Gaia-Sausage/Enceladus dwarf galaxy merger \citep[GSE;][]{belokurov2018, haywood2018, helmi2018}. This pair of stars has $|L_Z|$ values of $\approx$700\,kpc\,\kmsec \ that are slightly higher than most GSE stars near the Sun \citep[$|L_Z| \leq 500\,{\rm kpc\,\kmsec}$;][]{feuillet2020, limberg2022A, Carrillo2024gse}. However, tailored simulations do predict the GSE debris deposited in the Milky Way's outer halo to show higher $|L_Z|$ \citep{naidu2021, amarante2022}. Indeed, \citet{chandra2023gse} claimed to have identified such an outer-halo counterpart of GSE. Despite all of that, we caution that GSE occupies regions of kinematic/dynamical spaces expected to be crowded with substructures from other disrupted satellites, especially in this EMP regime \citep{yuan2020, limberg2021b, brauer2022}. \gui{Cosmological simulations show that the outer-halo stellar population of Milky Way-mass galaxies is expected to be dominated by the debris of a higher number of low-mass disrupted dwarfs in comparison to the inner halo (within the central $\sim$20\,kpc), which is composed of a small number of more massive mergers \citep[][see also \citealt{deason2016, deason2023}, \citealt{D'SouzaBell2018}, \citealt{brauer2019}, and \citealt{Monachesi2019aurigaHalos}]{fattahi2020}.}

Finally, \jone\, is on a moderate-eccentricity orbit ($e = 0.65^{+0.04}_{-0.06}$) with an acute $Y$-component of angular momentum ($L_Y = -4299^{+1415}_{-993}$\,kpc\,\kmsec). Such an exceptional $L_Y$ value is plausibly consistent with the aforementioned Sagittarius stream \citep{johnson2020sgr}, although that is not the case assuming a more restrictive set of criteria \citep{limberg2023}. In fact, \citet{ou2025sgr} has reported the existence of an EMP star in the core of Sagittarius dwarf spheroidal galaxy \citep{ibata1994} with similarly low Sr and Ba abundances. These authors interpreted this discovery as evidence for hierarchical assembly in Sagittarius since low levels of neutron-capture element abundances are a defining feature of ultra-faint dwarf galaxies \citep{ji2019ufds, simon2019}. We note that \jone\, does not spatially coincide with the main body of the Sagittarius stream \citep[e.g.,][]{vasiliev2019}, hence making this association tentative.

\section{Conclusions}
\label{conclusion}

In this work, we presented the first high-resolution spectroscopic follow-up of distant low-metallicity star candidates selected from the MAGIC survey, which includes five EMP stars and one UMP star. We confirm the accuracy of the MAGIC photometric metallicities and demonstrate that, even for carbon-enhanced stars, the success rate on identifying stars with \metal$\leq-3.0$ was nearly perfect. A comparison with literature data suggests that the program stars have chemistry consistent with the overall Milky Way halo population. The UMP star, \jtwo, has a light-element abundance pattern that can be explained by a Pop.~III progenitor with $\sim$11\Msun\, and low explosion energy. From a kinematical and dynamical analysis, we confirm that our sample stars belong to the distant outer halo population, with three stars having orbits that suggest, within uncertainties, \gui{they were either perturbed by the infall of the Magellanic system or were once members of the LMC that were later stripped by the Milky Way}. Finally, we speculate that the lower-than-average strontium and barium abundances for \jone\, could be explained by a membership in the Sagittarius stream\gui{, but origin in an ultra faint dwarf galaxy environment}. The addition of these six stars to the growing numbers of EMP and UMP stars in the literature helps further our understanding of the early stages in the chemical evolution of the Galaxy and the universe.

\begin{acknowledgments}

The work of V.M.P. is supported by NOIRLab, which is managed by the Association of Universities for Research in Astronomy (AURA) under a cooperative agreement with the U.S. National Science Foundation. G.L. acknowledges support from KICP/UChicago through a KICP Postdoctoral Fellowship. A.C. is supported by a Brinson Prize Fellowship at KICP/UChicago. S.R. acknowledges partial financial support from FAPESP (proc. 2020/15245-2), CNPq (proc. 303816/2022-8), and CAPES.
This research has made use of NASA's Astrophysics Data System Bibliographic
Services; the arXiv pre-print server operated by Cornell University; the
{\texttt{SIMBAD}} database hosted by the Strasbourg Astronomical Data Center;
and the online Q\&A platform {\texttt{stackoverflow}}
(\href{http://stackoverflow.com/}{http://stackoverflow.com/}).
The DELVE project is partially supported by the NASA Fermi Guest Investigator Program Cycle 9 No. 91201.
This work is partially supported by Fermilab LDRD project L2019-011. 
This material is based upon work supported by the National Science Foundation under Grant No. AST-2108168, AST-2108169, AST-2307126, and AST-2407526.
%
%
This project used data obtained with the Dark Energy Camera (DECam), which was constructed by the Dark Energy Survey (DES) collaboration. 
Funding for the DES Projects has been provided by the US Department of Energy, the U.S. National Science Foundation, the Ministry of Science and Education of Spain, the Science and Technology Facilities Council of the United Kingdom, the Higher Education Funding Council for England, the National Center for Supercomputing Applications at the University of Illinois at Urbana–Champaign, the Kavli Institute for Cosmological Physics at the University of Chicago, the Center for Cosmology and Astro-Particle Physics at the Ohio State University, the Mitchell Institute for Fundamental Physics and Astronomy at Texas A\&M University, Financiadora de Estudos e Projetos, Fundação Carlos Chagas Filho de Amparo à Pesquisa do Estado do Rio de Janeiro, Conselho 12 Nacional de Desenvolvimento Científico e Tecnológico and the Ministério da Ciência, Tecnologia e Inovação, the Deutsche Forschungsgemeinschaft and the Collaborating Institutions in the Dark Energy Survey.
The Collaborating Institutions are Argonne National Laboratory, the University of California at Santa Cruz, the University of Cambridge, Centro de Investigaciones Enérgeticas, Medioambientales y Tecnológicas–Madrid, the University of Chicago, University College London, the DES-Brazil Consortium, the University of Edinburgh, the Eidgenössische Technische Hochschule (ETH) Zürich, Fermi National Accelerator Laboratory, the University of Illinois at Urbana-Champaign, the Institut de Ciències de l’Espai (IEEC/CSIC), the Institut de Física d’Altes Energies, Lawrence Berkeley National Laboratory, the Ludwig-Maximilians Universität München and the associated Excellence Cluster Universe, the University of Michigan, NSF NOIRLab, the University of Nottingham, the Ohio State University, the OzDES Membership Consortium, the University of Pennsylvania, the University of Portsmouth, SLAC National Accelerator Laboratory, Stanford University, the University of Sussex, and Texas A\&M University.
Based on observations at NSF Cerro Tololo Inter-American Observatory, NSF NOIRLab (NOIRLab Prop. ID 2019A-0305; PI: Alex Drlica-Wagner, and NOIRLab Prop. ID 2023B-646244; PI: Anirudh Chiti), which is managed by the Association of Universities for Research in Astronomy (AURA) under a cooperative agreement with the U.S. National Science Foundation.
%
This manuscript has been authored by Fermi Research Alliance, LLC under Contract No. DE-AC02-07CH11359 with the U.S. Department of Energy, Office of Science, Office of High Energy Physics. The United States Government retains and the publisher, by accepting the article for publication, acknowledges that the United States Government retains a non-exclusive, paid-up, irrevocable, world-wide license to publish or reproduce the published form of this manuscript, or allow others to do so, for United States Government purposes.
This work has made use of data from the European Space Agency (ESA) mission Gaia (\url{https://www.cosmos.esa.int/gaia}), processed by the Gaia Data Processing and Analysis Consortium (DPAC, \url{https://www.cosmos.esa.int/web/gaia/dpac/consortium}). Funding for the DPAC has been provided by national institutions, in particular the institutions participating in the Gaia Multilateral Agreement.

\end{acknowledgments}

\software{
{\texttt{astropy}}\,\citep{astropy2013,astropy2018}, 
{\texttt{awk}}\,\citep{awk}, 
{\texttt{dustmaps}}\,\citep{green2018}, 
{\texttt{gnuplot}}\,\citep{gnuplot}, 
{\texttt{NOIRLab IRAF}}\,\citep{tody1986,tody1993,fitzpatrick2024}, 
{\texttt{linemake}}\,\citep{placco2021,placco2021a},
{\texttt{matplotlib}}\,\citep{matplotlib}, 
{\texttt{MOOG}}\,\citep{sneden1973},  
{\texttt{numpy}}\,\citep{numpy}, 
{\texttt{pandas}}\,\citep{pandas}, 
{\texttt{RVSearch}}\,\citep{Rosenthal2021}, 
{\texttt{sed}}\,\citep{sed},
{\texttt{stilts}}\,\citep{stilts}.
}

\facilities{Blanco (DECam), Magellan:Clay (MIKE)}

\vfill
\newpage

\bibliographystyle{aasjournal}
\bibliography{bibliografia}

\twocolumngrid

\startlongtable

\appendix
\restartappendixnumbering

\section{Atomic line list}

\begin{deluxetable}{lrrrrrrrrrrrrrrr} 
\tabletypesize{\tiny}
\tablewidth{0pc}
\tablecaption{\label{eqwl} Atomic Data and Derived Abundances}
\tablehead{
\colhead{}&
\colhead{}&
\colhead{}&
\colhead{}&
\multicolumn{2}{c}{\jone}&
\multicolumn{2}{c}{\jtwo}&
\multicolumn{2}{c}{\jthree}&
\multicolumn{2}{c}{\jfour}&
\multicolumn{2}{c}{\jfive}&
\multicolumn{2}{c}{\jsix}\\
\colhead{Ion}&
\colhead{$\lambda$}&
\colhead{$\chi$} &
\colhead{$\log\,gf$}&
\colhead{$EW$}&
\colhead{$\log\epsilon$\,(X)}&
\colhead{$EW$}&
\colhead{$\log\epsilon$\,(X)}&
\colhead{$EW$}&
\colhead{$\log\epsilon$\,(X)}&
\colhead{$EW$}&
\colhead{$\log\epsilon$\,(X)}&
\colhead{$EW$}&
\colhead{$\log\epsilon$\,(X)}&
\colhead{$EW$}&
\colhead{$\log\epsilon$\,(X)}\\
\colhead{}&
\colhead{({\AA})}&
\colhead{(eV)} &
\colhead{}&
\colhead{(m{\AA})}&
\colhead{}&
\colhead{(m{\AA})}&
\colhead{}&
\colhead{(m{\AA})}&
\colhead{}&
\colhead{(m{\AA})}&
\colhead{}&
\colhead{(m{\AA})}&
\colhead{}&
\colhead{(m{\AA})}&
\colhead{}}
\startdata
         CH & 4303.00 & \nodata & \nodata  &     syn &     4.62  &     syn &     6.01 &     syn &     5.23 &     syn &     4.68 &     syn &     5.14 &     syn &     4.56  \\
         CH & 4313.00 & \nodata & \nodata  &     syn &     4.62  &     syn &     6.01 &     syn &     5.23 &     syn &     4.68 &     syn &     5.04 &     syn &     4.56  \\
\ion{Na}{1} & 5889.95 & 0.00 &    0.11 &   155.27 &     3.16 &    45.55 &     1.97 &  \nodata &  \nodata &  \nodata &  \nodata &  \nodata &  \nodata &   155.97 &     3.16 \\
\ion{Na}{1} & 5895.92 & 0.00 & $-$0.19 &   128.07 &     3.01 &    38.38 &     2.16 &   112.68 &     3.08 &   177.15 &     3.86 &   120.86 &     3.17 &   130.73 &     3.02 \\
\ion{Mg}{1} & 4167.27 & 4.35 & $-$0.74 &  \nodata &  \nodata &  \nodata &  \nodata &  \nodata &  \nodata &  \nodata &  \nodata &  \nodata &  \nodata &    35.86 &     4.85 \\
\ion{Mg}{1} & 4702.99 & 4.33 & $-$0.44 &  \nodata &  \nodata &  \nodata &  \nodata &    42.05 &     4.73 &    69.64 &     5.03 &    31.28 &     4.51 &    49.90 &     4.73 \\
\ion{Mg}{1} & 5172.68 & 2.71 & $-$0.36 &   184.80 &     4.76 &    69.59 &     3.40 &   155.57 &     4.59 &   199.52 &     4.90 &   144.46 &     4.38 &   169.88 &     4.54 \\
\ion{Mg}{1} & 5183.60 & 2.72 & $-$0.17 &   196.52 &     4.71 &    92.00 &     3.60 &   176.89 &     4.67 &   222.33 &     4.91 &   175.23 &     4.62 &   213.58 &     4.86 \\
\ion{Mg}{1} & 5528.40 & 4.35 & $-$0.55 &    40.55 &     4.70 &  \nodata &  \nodata &  \nodata &  \nodata &    46.76 &     4.79 &  \nodata &  \nodata &    57.00 &     4.93 \\
\ion{Al}{1} & 3944.00 & 0.00  &  $-$0.64  &     syn &     2.24  & \nodata &  \nodata &     syn &     2.50 & \nodata &     2.90 &     syn &     2.11 &     syn &     2.30  \\
\ion{Al}{1} & 3961.52 & 0.01  &  $-$0.33  &     syn &     2.34  &     syn &     1.45 &     syn &     2.40 & \nodata &     2.90 &     syn &     2.18 &     syn &     2.40  \\
\ion{Si}{1} & 3905.52 & 1.91  &  $-$1.04  &     syn &     4.55  & \nodata &  \nodata &     syn &     4.78 &     syn &     4.36 &     syn &     4.67 &     syn &     4.66  \\
\ion{Si}{1} & 4102.94 & 1.91  &  $-$3.34  &     syn &     4.62  & \nodata &  \nodata &     syn &     4.76 & \nodata &     4.46 & \nodata &  \nodata &     syn &     4.61  \\
\ion{Ca}{1} & 4226.74 & 0.00 &    0.24 &  \nodata &  \nodata &    96.20 &     2.28 &  \nodata &  \nodata &  \nodata &  \nodata &   181.38 &     3.28 &  \nodata &  \nodata \\
\ion{Ca}{1} & 4283.01 & 1.89 & $-$0.20 &    40.64 &     3.35 &  \nodata &  \nodata &  \nodata &  \nodata &    59.56 &     3.67 &  \nodata &  \nodata &  \nodata &  \nodata \\
\ion{Ca}{1} & 4318.65 & 1.90 & $-$0.21 &    42.56 &     3.40 &  \nodata &  \nodata &  \nodata &  \nodata &  \nodata &  \nodata &  \nodata &  \nodata &    31.65 &     3.17 \\
\ion{Ca}{1} & 4425.44 & 1.88 & $-$0.41 &    37.64 &     3.48 &  \nodata &  \nodata &  \nodata &  \nodata &    35.01 &     3.43 &  \nodata &  \nodata &  \nodata &  \nodata \\
\ion{Ca}{1} & 4454.78 & 1.90 &    0.26 &  \nodata &  \nodata &  \nodata &  \nodata &  \nodata &  \nodata &  \nodata &  \nodata &    59.59 &     3.33 &    57.60 &     3.14 \\
\ion{Ca}{1} & 4455.89 & 1.90 & $-$0.55 &  \nodata &  \nodata &  \nodata &  \nodata &    20.90 &     3.45 &    34.77 &     3.58 &  \nodata &  \nodata &    23.48 &     3.33 \\
\ion{Ca}{1} & 5588.76 & 2.52 &    0.30 &    28.13 &     3.31 &  \nodata &  \nodata &    30.94 &     3.50 &    41.26 &     3.54 &  \nodata &  \nodata &  \nodata &  \nodata \\
\ion{Ca}{1} & 6102.72 & 1.88 & $-$0.81 &    24.00 &     3.52 &  \nodata &  \nodata &  \nodata &  \nodata &    37.04 &     3.76 &  \nodata &  \nodata &  \nodata &  \nodata \\
\ion{Ca}{1} & 6122.22 & 1.89 & $-$0.33 &  \nodata &  \nodata &  \nodata &  \nodata &    42.21 &     3.56 &    61.21 &     3.67 &    42.42 &     3.51 &    44.55 &     3.39 \\
\ion{Ca}{1} & 6162.17 & 1.90 & $-$0.11 &    65.24 &     3.52 &  \nodata &  \nodata &    50.42 &     3.49 &    71.74 &     3.62 &    61.03 &     3.60 &    60.78 &     3.42 \\
\ion{Ca}{1} & 6439.07 & 2.52 &    0.33 &    36.71 &     3.40 &  \nodata &  \nodata &  \nodata &  \nodata &    44.11 &     3.52 &    33.20 &     3.44 &    41.65 &     3.45 \\
\ion{Sc}{2} & 4246.82 & 0.32 &    0.24 &   117.19 &  $-$0.23 &    59.74 &  $-$0.64 &   108.10 &     0.15 &  \nodata &  \nodata &    99.98 &  $-$0.19 &  \nodata &  \nodata \\
\ion{Sc}{2} & 4320.73 & 0.60 & $-$0.25 &    80.03 &  $-$0.17 &  \nodata &  \nodata &    70.73 &     0.11 &  \nodata &  \nodata &  \nodata &  \nodata &  \nodata &  \nodata \\
\ion{Sc}{2} & 4324.98 & 0.59 & $-$0.44 &    61.28 &  $-$0.31 &  \nodata &  \nodata &    59.04 &     0.07 &  \nodata &  \nodata &  \nodata &  \nodata &    60.58 &  $-$0.33 \\
\ion{Sc}{2} & 4374.45 & 0.62 & $-$0.42 &    63.19 &  $-$0.27 &  \nodata &  \nodata &    55.25 &     0.01 &  \nodata &  \nodata &    49.58 &  $-$0.20 &    61.65 &  $-$0.30 \\
\ion{Sc}{2} & 4400.39 & 0.60 & $-$0.54 &    54.76 &  $-$0.30 &  \nodata &  \nodata &    54.65 &     0.10 &    65.36 &  $-$0.04 &    47.78 &  $-$0.14 &  \nodata &  \nodata \\
\ion{Sc}{2} & 4415.54 & 0.59 & $-$0.67 &    49.15 &  $-$0.28 &  \nodata &  \nodata &  \nodata &  \nodata &  \nodata &  \nodata &    52.35 &     0.05 &    56.54 &  $-$0.18 \\
\ion{Ti}{1} & 3998.64 & 0.05 &    0.02 &  \nodata &  \nodata &    13.56 &     1.35 &  \nodata &  \nodata &    58.92 &     1.68 &    48.58 &     1.72 &  \nodata &  \nodata \\
\ion{Ti}{1} & 4533.24 & 0.85 &    0.54 &  \nodata &  \nodata &  \nodata &  \nodata &  \nodata &  \nodata &  \nodata &  \nodata &    37.29 &     1.91 &    34.19 &     1.63 \\
\ion{Ti}{1} & 4534.78 & 0.84 &    0.35 &  \nodata &  \nodata &  \nodata &  \nodata &    31.51 &     2.06 &  \nodata &  \nodata &  \nodata &  \nodata &    23.46 &     1.59 \\
\ion{Ti}{1} & 4681.91 & 0.05 & $-$1.01 &    14.92 &     1.76 &  \nodata &  \nodata &    15.13 &     2.06 &    27.92 &     2.08 &  \nodata &  \nodata &  \nodata &  \nodata \\
\ion{Ti}{1} & 4981.73 & 0.84 &    0.57 &  \nodata &  \nodata &  \nodata &  \nodata &  \nodata &  \nodata &  \nodata &  \nodata &    34.91 &     1.79 &    44.96 &     1.73 \\
\ion{Ti}{1} & 4991.07 & 0.84 &    0.45 &    39.33 &     1.81 &  \nodata &  \nodata &    49.04 &     2.23 &    54.99 &     2.05 &  \nodata &  \nodata &  \nodata &  \nodata \\
\ion{Ti}{1} & 4999.50 & 0.83 &    0.32 &    34.37 &     1.84 &  \nodata &  \nodata &    33.61 &     2.08 &    50.24 &     2.10 &  \nodata &  \nodata &    31.18 &     1.73 \\
\ion{Ti}{2} & 3913.46 & 1.12 & $-$0.36 &  \nodata &  \nodata &  \nodata &  \nodata &  \nodata &  \nodata &  \nodata &  \nodata &  \nodata &  \nodata &   114.33 &     1.92 \\
\ion{Ti}{2} & 4337.91 & 1.08 & $-$0.96 &  \nodata &  \nodata &  \nodata &  \nodata &    89.86 &     2.20 &  \nodata &  \nodata &    77.30 &     1.80 &    91.96 &     1.77 \\
\ion{Ti}{2} & 4394.06 & 1.22 & $-$1.77 &    47.10 &     1.97 &  \nodata &  \nodata &  \nodata &  \nodata &  \nodata &  \nodata &  \nodata &  \nodata &    49.47 &     2.00 \\
\ion{Ti}{2} & 4395.03 & 1.08 & $-$0.54 &   120.99 &     1.95 &  \nodata &  \nodata &   117.52 &     2.46 &  \nodata &  \nodata &   105.11 &     2.01 &  \nodata &  \nodata \\
\ion{Ti}{2} & 4395.84 & 1.24 & $-$1.93 &    30.49 &     1.87 &  \nodata &  \nodata &  \nodata &  \nodata &  \nodata &  \nodata &  \nodata &  \nodata &  \nodata &  \nodata \\
\ion{Ti}{2} & 4399.77 & 1.24 & $-$1.20 &    69.63 &     1.79 &  \nodata &  \nodata &    80.49 &     2.41 &    88.91 &     2.27 &    53.51 &     1.79 &    79.52 &     1.96 \\
\ion{Ti}{2} & 4417.71 & 1.17 & $-$1.19 &  \nodata &  \nodata &  \nodata &  \nodata &  \nodata &  \nodata &    73.05 &     1.86 &    54.79 &     1.71 &    88.91 &     2.04 \\
\ion{Ti}{2} & 4418.33 & 1.24 & $-$1.99 &  \nodata &  \nodata &  \nodata &  \nodata &  \nodata &  \nodata &    35.99 &     2.12 &  \nodata &  \nodata &    33.86 &     1.98 \\
\ion{Ti}{2} & 4443.80 & 1.08 & $-$0.71 &   116.82 &     2.02 &    38.19 &     1.17 &   107.71 &     2.36 &  \nodata &  \nodata &    96.34 &     1.95 &   111.64 &     1.92 \\
\ion{Ti}{2} & 4444.55 & 1.12 & $-$2.20 &  \nodata &  \nodata &  \nodata &  \nodata &    28.51 &     2.30 &    33.97 &     2.14 &  \nodata &  \nodata &    32.73 &     2.01 \\
\ion{Ti}{2} & 4450.48 & 1.08 & $-$1.52 &    73.97 &     1.98 &    11.23 &     1.29 &    67.79 &     2.27 &    81.10 &     2.22 &    43.90 &     1.74 &  \nodata &  \nodata \\
\ion{Ti}{2} & 4464.45 & 1.16 & $-$1.81 &    46.49 &     1.92 &  \nodata &  \nodata &    48.34 &     2.31 &    54.34 &     2.14 &  \nodata &  \nodata &  \nodata &  \nodata \\
\ion{Ti}{2} & 4470.85 & 1.17 & $-$2.02 &    31.53 &     1.88 &  \nodata &  \nodata &    25.43 &     2.11 &  \nodata &  \nodata &  \nodata &  \nodata &    38.40 &     2.00 \\
\ion{Ti}{2} & 4501.27 & 1.12 & $-$0.77 &   102.13 &     1.78 &  \nodata &  \nodata &  \nodata &  \nodata &   106.85 &     2.06 &  \nodata &  \nodata &   113.93 &     2.06 \\
\ion{Ti}{2} & 4533.96 & 1.24 & $-$0.53 &  \nodata &  \nodata &    41.96 &     1.23 &    98.68 &     2.12 &  \nodata &  \nodata &  \nodata &  \nodata &   122.06 &     2.13 \\
\ion{Ti}{2} & 4571.97 & 1.57 & $-$0.31 &    92.33 &     1.70 &  \nodata &  \nodata &  \nodata &  \nodata &  \nodata &  \nodata &  \nodata &  \nodata &   108.31 &     2.04 \\
\ion{Ti}{2} & 4708.66 & 1.24 & $-$2.35 &  \nodata &  \nodata &  \nodata &  \nodata &  \nodata &  \nodata &    24.19 &     2.21 &  \nodata &  \nodata &  \nodata &  \nodata \\
\ion{Ti}{2} & 4798.53 & 1.08 & $-$2.66 &    14.69 &     1.97 &  \nodata &  \nodata &  \nodata &  \nodata &  \nodata &  \nodata &  \nodata &  \nodata &  \nodata &  \nodata \\
\ion{Ti}{2} & 5129.16 & 1.89 & $-$1.34 &  \nodata &  \nodata &  \nodata &  \nodata &  \nodata &  \nodata &  \nodata &  \nodata &  \nodata &  \nodata &    33.25 &     2.08 \\
\ion{Ti}{2} & 5188.69 & 1.58 & $-$1.05 &    64.04 &     1.88 &  \nodata &  \nodata &  \nodata &  \nodata &  \nodata &  \nodata &    45.92 &     1.85 &  \nodata &  \nodata \\
\ion{Ti}{2} & 5336.79 & 1.58 & $-$1.60 &  \nodata &  \nodata &  \nodata &  \nodata &  \nodata &  \nodata &  \nodata &  \nodata &  \nodata &  \nodata &    38.94 &     2.03 \\
\ion{Ti}{2} & 5381.02 & 1.57 & $-$1.97 &  \nodata &  \nodata &  \nodata &  \nodata &  \nodata &  \nodata &  \nodata &  \nodata &  \nodata &  \nodata &    20.02 &     2.01 \\
\ion{Cr}{1} & 4274.80 & 0.00 & $-$0.22 &  \nodata &  \nodata &  \nodata &  \nodata &    95.01 &     2.14 &  \nodata &  \nodata &  \nodata &  \nodata &  \nodata &  \nodata \\
\ion{Cr}{1} & 4289.72 & 0.00 & $-$0.37 &   111.73 &     2.22 &    29.45 &     1.27 &    91.66 &     2.20 &   105.12 &     2.14 &  \nodata &  \nodata &  \nodata &  \nodata \\
\ion{Cr}{1} & 4646.15 & 1.03 & $-$0.74 &    28.60 &     2.26 &  \nodata &  \nodata &  \nodata &  \nodata &    24.97 &     2.17 &  \nodata &  \nodata &  \nodata &  \nodata \\
\ion{Cr}{1} & 5206.04 & 0.94 &    0.02 &  \nodata &  \nodata &  \nodata &  \nodata &    58.50 &     2.12 &    61.65 &     1.89 &    60.19 &     2.07 &    79.81 &     2.13 \\
\ion{Cr}{1} & 5298.28 & 0.98 & $-$1.14 &    18.07 &     2.31 &  \nodata &  \nodata &  \nodata &  \nodata &  \nodata &  \nodata &  \nodata &  \nodata &  \nodata &  \nodata \\
\ion{Cr}{1} & 5409.77 & 1.03 & $-$0.67 &    40.83 &     2.35 &  \nodata &  \nodata &  \nodata &  \nodata &    33.55 &     2.22 &  \nodata &  \nodata &    25.03 &     2.01 \\
\ion{Mn}{1} & 4030.75 &    0.00  &  $-$0.50  &     syn &     1.27  & \nodata &  \nodata &     syn &     1.61 &     syn &     1.18 &     syn &     1.19 &     syn &     1.33  \\
\ion{Mn}{1} & 4033.06 &    0.00  &  $-$0.65  &     syn &     1.27  & \nodata &  \nodata &     syn &     1.61 &     syn &     1.18 &     syn &     1.19 &     syn &     1.33  \\
\ion{Mn}{1} & 4034.48 &    0.00  &  $-$0.84  &     syn &     1.27  & \nodata &  \nodata &     syn &     1.61 &     syn &     1.18 &     syn &     1.19 &     syn &     1.33  \\
\ion{Fe}{1} & 3763.79 & 0.99 & $-$0.22 &  \nodata &  \nodata &    95.31 &     3.40 &  \nodata &  \nodata &  \nodata &  \nodata &  \nodata &  \nodata &  \nodata &  \nodata \\
\ion{Fe}{1} & 3787.88 & 1.01 & $-$0.84 &   161.46 &     4.56 &    62.78 &     3.14 &  \nodata &  \nodata &  \nodata &  \nodata &  \nodata &  \nodata &  \nodata &  \nodata \\
\ion{Fe}{1} & 3815.84 & 1.48 &    0.24 &  \nodata &  \nodata &    93.78 &     3.42 &  \nodata &  \nodata &  \nodata &  \nodata &  \nodata &  \nodata &  \nodata &  \nodata \\
\ion{Fe}{1} & 3820.43 & 0.86 &    0.16 &  \nodata &  \nodata &   111.21 &     3.27 &  \nodata &  \nodata &  \nodata &  \nodata &  \nodata &  \nodata &  \nodata &  \nodata \\
\ion{Fe}{1} & 3825.88 & 0.91 & $-$0.02 &  \nodata &  \nodata &   104.95 &     3.34 &  \nodata &  \nodata &  \nodata &  \nodata &  \nodata &  \nodata &  \nodata &  \nodata \\
\ion{Fe}{1} & 3827.82 & 1.56 &    0.09 &  \nodata &  \nodata &    89.17 &     3.52 &  \nodata &  \nodata &  \nodata &  \nodata &  \nodata &  \nodata &  \nodata &  \nodata \\
\ion{Fe}{1} & 3840.44 & 0.99 & $-$0.50 &   182.37 &     4.37 &    79.16 &     3.17 &  \nodata &  \nodata &  \nodata &  \nodata &  \nodata &  \nodata &  \nodata &  \nodata \\
\ion{Fe}{1} & 3841.05 & 1.61 & $-$0.04 &  \nodata &  \nodata &  \nodata &  \nodata &  \nodata &  \nodata &  \nodata &  \nodata &  \nodata &  \nodata &   139.03 &     4.08 \\
\ion{Fe}{1} & 3849.97 & 1.01 & $-$0.86 &   178.07 &     4.70 &    72.41 &     3.38 &  \nodata &  \nodata &  \nodata &  \nodata &  \nodata &  \nodata &   152.67 &     4.34 \\
\ion{Fe}{1} & 3856.37 & 0.05 & $-$1.28 &   205.27 &     4.57 &  \nodata &  \nodata &  \nodata &  \nodata &  \nodata &  \nodata &  \nodata &  \nodata &  \nodata &  \nodata \\
\ion{Fe}{1} & 3865.52 & 1.01 & $-$0.95 &  \nodata &  \nodata &    75.43 &     3.54 &   127.10 &     4.44 &   152.63 &     4.50 &   126.15 &     4.32 &  \nodata &  \nodata \\
\ion{Fe}{1} & 3878.02 & 0.96 & $-$0.90 &  \nodata &  \nodata &    70.50 &     3.30 &  \nodata &  \nodata &  \nodata &  \nodata &   129.41 &     4.28 &   159.30 &     4.41 \\
\ion{Fe}{1} & 3899.71 & 0.09 & $-$1.52 &   184.60 &     4.57 &    81.71 &     3.23 &   136.90 &     4.25 &  \nodata &  \nodata &  \nodata &  \nodata &  \nodata &  \nodata \\
\ion{Fe}{1} & 3902.95 & 1.56 & $-$0.44 &   151.21 &     4.62 &    70.59 &     3.53 &  \nodata &  \nodata &  \nodata &  \nodata &  \nodata &  \nodata &   129.25 &     4.18 \\
\ion{Fe}{1} & 3920.26 & 0.12 & $-$1.73 &   165.22 &     4.49 &  \nodata &  \nodata &   126.93 &     4.26 &  \nodata &  \nodata &  \nodata &  \nodata &  \nodata &  \nodata \\
\ion{Fe}{1} & 3949.95 & 2.18 & $-$1.25 &    60.18 &     4.21 &  \nodata &  \nodata &    66.98 &     4.65 &    61.20 &     4.25 &  \nodata &  \nodata &    62.76 &     4.22 \\
\ion{Fe}{1} & 3977.74 & 2.20 & $-$1.12 &    63.79 &     4.18 &  \nodata &  \nodata &  \nodata &  \nodata &  \nodata &  \nodata &  \nodata &  \nodata &    55.43 &     3.98 \\
\ion{Fe}{1} & 4005.24 & 1.56 & $-$0.58 &   137.66 &     4.49 &    50.74 &     3.22 &   122.38 &     4.59 &   117.60 &     4.12 &    97.14 &     3.87 &   130.13 &     4.31 \\
\ion{Fe}{1} & 4045.81 & 1.49 &    0.28 &  \nodata &  \nodata &    94.40 &     3.33 &  \nodata &  \nodata &  \nodata &  \nodata &   141.68 &     3.93 &  \nodata &  \nodata \\
\ion{Fe}{1} & 4063.59 & 1.56 &    0.06 &   165.46 &     4.30 &    89.22 &     3.48 &  \nodata &  \nodata &  \nodata &  \nodata &   129.78 &     4.00 &   169.65 &     4.34 \\
\ion{Fe}{1} & 4067.98 & 3.21 & $-$0.53 &    36.08 &     4.32 &  \nodata &  \nodata &  \nodata &  \nodata &  \nodata &  \nodata &  \nodata &  \nodata &    22.51 &     4.00 \\
\ion{Fe}{1} & 4071.74 & 1.61 & $-$0.01 &  \nodata &  \nodata &    72.73 &     3.17 &  \nodata &  \nodata &  \nodata &  \nodata &   125.97 &     4.05 &   149.39 &     4.16 \\
\ion{Fe}{1} & 4132.06 & 1.61 & $-$0.68 &   142.03 &     4.67 &    40.42 &     3.16 &   112.95 &     4.48 &   132.63 &     4.56 &    98.31 &     4.01 &  \nodata &  \nodata \\
\ion{Fe}{1} & 4134.68 & 2.83 & $-$0.65 &  \nodata &  \nodata &  \nodata &  \nodata &  \nodata &  \nodata &    46.02 &     4.15 &  \nodata &  \nodata &  \nodata &  \nodata \\
\ion{Fe}{1} & 4143.87 & 1.56 & $-$0.51 &   139.36 &     4.38 &    57.96 &     3.27 &  \nodata &  \nodata &   133.53 &     4.34 &   108.45 &     4.04 &   143.29 &     4.45 \\
\ion{Fe}{1} & 4147.67 & 1.49 & $-$2.07 &    78.41 &     4.48 &  \nodata &  \nodata &    58.67 &     4.44 &    66.90 &     4.28 &    60.86 &     4.39 &    73.92 &     4.35 \\
\ion{Fe}{1} & 4154.50 & 2.83 & $-$0.69 &    47.46 &     4.22 &  \nodata &  \nodata &    50.92 &     4.51 &  \nodata &  \nodata &  \nodata &  \nodata &    32.17 &     3.90 \\
\ion{Fe}{1} & 4156.80 & 2.83 & $-$0.81 &    42.09 &     4.24 &  \nodata &  \nodata &  \nodata &  \nodata &  \nodata &  \nodata &  \nodata &  \nodata &    31.31 &     4.00 \\
\ion{Fe}{1} & 4157.78 & 3.42 & $-$0.40 &    37.12 &     4.47 &  \nodata &  \nodata &  \nodata &  \nodata &  \nodata &  \nodata &  \nodata &  \nodata &  \nodata &  \nodata \\
\ion{Fe}{1} & 4158.79 & 3.43 & $-$0.70 &    26.72 &     4.57 &  \nodata &  \nodata &  \nodata &  \nodata &  \nodata &  \nodata &  \nodata &  \nodata &  \nodata &  \nodata \\
\ion{Fe}{1} & 4174.91 & 0.91 & $-$2.94 &    70.19 &     4.46 &  \nodata &  \nodata &  \nodata &  \nodata &  \nodata &  \nodata &    32.99 &     4.03 &    68.63 &     4.38 \\
\ion{Fe}{1} & 4175.64 & 2.85 & $-$0.83 &    38.91 &     4.23 &  \nodata &  \nodata &    42.78 &     4.52 &  \nodata &  \nodata &  \nodata &  \nodata &    25.29 &     3.92 \\
\ion{Fe}{1} & 4181.76 & 2.83 & $-$0.37 &    59.39 &     4.10 &  \nodata &  \nodata &    55.44 &     4.27 &    62.79 &     4.17 &  \nodata &  \nodata &    55.52 &     3.99 \\
\ion{Fe}{1} & 4182.38 & 3.02 & $-$1.18 &    16.30 &     4.28 &  \nodata &  \nodata &  \nodata &  \nodata &  \nodata &  \nodata &  \nodata &  \nodata &  \nodata &  \nodata \\
\ion{Fe}{1} & 4184.89 & 2.83 & $-$0.87 &    38.38 &     4.23 &  \nodata &  \nodata &    27.22 &     4.23 &  \nodata &  \nodata &  \nodata &  \nodata &    26.14 &     3.95 \\
\ion{Fe}{1} & 4187.04 & 2.45 & $-$0.56 &    94.40 &     4.50 &    16.22 &     3.43 &    70.82 &     4.30 &    76.38 &     4.14 &    66.24 &     4.13 &    70.17 &     3.97 \\
\ion{Fe}{1} & 4187.80 & 2.42 & $-$0.51 &   100.81 &     4.55 &  \nodata &  \nodata &    87.97 &     4.60 &  \nodata &  \nodata &    60.82 &     3.94 &    73.68 &     3.95 \\
\ion{Fe}{1} & 4191.43 & 2.47 & $-$0.67 &  \nodata &  \nodata &  \nodata &  \nodata &    60.88 &     4.24 &    76.90 &     4.29 &  \nodata &  \nodata &    61.38 &     3.94 \\
\ion{Fe}{1} & 4195.33 & 3.33 & $-$0.49 &    40.44 &     4.50 &  \nodata &  \nodata &  \nodata &  \nodata &  \nodata &  \nodata &  \nodata &  \nodata &    24.02 &     4.14 \\
\ion{Fe}{1} & 4199.10 & 3.05 &    0.16 &  \nodata &  \nodata &  \nodata &  \nodata &  \nodata &  \nodata &  \nodata &  \nodata &  \nodata &  \nodata &    73.80 &     4.08 \\
\ion{Fe}{1} & 4202.03 & 1.49 & $-$0.69 &   146.50 &     4.60 &    52.66 &     3.26 &  \nodata &  \nodata &  \nodata &  \nodata &   102.18 &     3.96 &   124.94 &     4.13 \\
\ion{Fe}{1} & 4216.18 & 0.00 & $-$3.36 &   118.44 &     4.75 &  \nodata &  \nodata &  \nodata &  \nodata &   108.28 &     4.61 &    88.50 &     4.45 &  \nodata &  \nodata \\
\ion{Fe}{1} & 4222.21 & 2.45 & $-$0.91 &    55.72 &     4.10 &  \nodata &  \nodata &    65.51 &     4.54 &    52.30 &     4.04 &    36.64 &     3.93 &    55.17 &     4.05 \\
\ion{Fe}{1} & 4227.43 & 3.33 &    0.27 &    68.69 &     4.24 &  \nodata &  \nodata &  \nodata &  \nodata &    62.78 &     4.14 &    60.62 &     4.25 &    73.63 &     4.30 \\
\ion{Fe}{1} & 4233.60 & 2.48 & $-$0.60 &    85.96 &     4.38 &  \nodata &  \nodata &    63.02 &     4.21 &    80.76 &     4.30 &    59.37 &     4.07 &  \nodata &  \nodata \\
\ion{Fe}{1} & 4238.81 & 3.40 & $-$0.23 &    51.38 &     4.51 &  \nodata &  \nodata &    27.86 &     4.27 &  \nodata &  \nodata &  \nodata &  \nodata &    27.72 &     4.04 \\
\ion{Fe}{1} & 4250.12 & 2.47 & $-$0.38 &    87.77 &     4.18 &    20.43 &     3.39 &    79.75 &     4.33 &  \nodata &  \nodata &    74.91 &     4.14 &    81.02 &     4.01 \\
\ion{Fe}{1} & 4250.79 & 1.56 & $-$0.71 &   139.66 &     4.55 &    46.25 &     3.23 &   107.39 &     4.28 &  \nodata &  \nodata &   100.96 &     4.01 &   124.81 &     4.21 \\
\ion{Fe}{1} & 4260.47 & 2.40 &    0.08 &  \nodata &  \nodata &    35.05 &     3.18 &   101.81 &     4.28 &  \nodata &  \nodata &    95.33 &     4.06 &   119.63 &     4.31 \\
\ion{Fe}{1} & 4271.15 & 2.45 & $-$0.34 &    93.15 &     4.23 &    27.40 &     3.50 &    82.70 &     4.32 &    88.48 &     4.16 &    81.11 &     4.21 &    87.83 &     4.08 \\
\ion{Fe}{1} & 4271.76 & 1.49 & $-$0.17 &   161.83 &     4.31 &    76.24 &     3.24 &   133.49 &     4.23 &   161.36 &     4.36 &   131.21 &     4.10 &   163.88 &     4.33 \\
\ion{Fe}{1} & 4282.40 & 2.18 & $-$0.78 &   105.63 &     4.64 &  \nodata &  \nodata &    78.09 &     4.36 &    99.18 &     4.55 &    66.67 &     4.03 &  \nodata &  \nodata \\
\ion{Fe}{1} & 4325.76 & 1.61 &    0.01 &   169.47 &     4.39 &    75.57 &     3.17 &   141.34 &     4.33 &   165.07 &     4.37 &   133.53 &     4.10 &   157.72 &     4.19 \\
\ion{Fe}{1} & 4352.73 & 2.22 & $-$1.29 &    66.56 &     4.36 &  \nodata &  \nodata &    60.22 &     4.54 &    52.45 &     4.13 &  \nodata &  \nodata &    63.73 &     4.27 \\
\ion{Fe}{1} & 4375.93 & 0.00 & $-$3.00 &   135.19 &     4.70 &    41.47 &     3.62 &    93.60 &     4.29 &   108.44 &     4.19 &    89.65 &     4.06 &  \nodata &  \nodata \\
\ion{Fe}{1} & 4404.75 & 1.56 & $-$0.15 &   177.02 &     4.53 &    77.35 &     3.30 &   133.15 &     4.24 &   145.80 &     4.12 &   141.45 &     4.31 &   154.55 &     4.18 \\
\ion{Fe}{1} & 4415.12 & 1.61 & $-$0.62 &   136.28 &     4.37 &    58.46 &     3.41 &  \nodata &  \nodata &   121.42 &     4.14 &   117.63 &     4.34 &   136.13 &     4.35 \\
\ion{Fe}{1} & 4427.31 & 0.05 & $-$2.92 &   141.55 &     4.80 &    45.39 &     3.66 &    92.82 &     4.24 &   122.56 &     4.51 &    96.68 &     4.21 &  \nodata &  \nodata \\
\ion{Fe}{1} & 4430.61 & 2.22 & $-$1.73 &    43.75 &     4.41 &  \nodata &  \nodata &    29.71 &     4.40 &    43.73 &     4.41 &  \nodata &  \nodata &    43.92 &     4.37 \\
\ion{Fe}{1} & 4442.34 & 2.20 & $-$1.23 &    81.10 &     4.52 &  \nodata &  \nodata &    59.39 &     4.41 &  \nodata &  \nodata &    61.93 &     4.38 &    56.70 &     4.05 \\
\ion{Fe}{1} & 4443.19 & 2.86 & $-$1.04 &    31.83 &     4.30 &  \nodata &  \nodata &    32.83 &     4.53 &  \nodata &  \nodata &    24.34 &     4.28 &    20.76 &     4.01 \\
\ion{Fe}{1} & 4447.72 & 2.22 & $-$1.36 &    63.09 &     4.35 &  \nodata &  \nodata &    51.59 &     4.43 &    57.82 &     4.27 &    47.65 &     4.28 &    46.79 &     4.04 \\
\ion{Fe}{1} & 4461.65 & 0.09 & $-$3.19 &   129.32 &     4.82 &    30.68 &     3.69 &    87.05 &     4.40 &   114.04 &     4.60 &  \nodata &  \nodata &  \nodata &  \nodata \\
\ion{Fe}{1} & 4466.55 & 2.83 & $-$0.60 &  \nodata &  \nodata &  \nodata &  \nodata &    66.29 &     4.67 &    68.65 &     4.47 &  \nodata &  \nodata &  \nodata &  \nodata \\
\ion{Fe}{1} & 4489.74 & 0.12 & $-$3.90 &  \nodata &  \nodata &  \nodata &  \nodata &    51.78 &     4.43 &    79.97 &     4.56 &    57.67 &     4.42 &  \nodata &  \nodata \\
\ion{Fe}{1} & 4494.56 & 2.20 & $-$1.14 &    88.87 &     4.56 &  \nodata &  \nodata &    57.22 &     4.28 &    66.38 &     4.17 &    63.02 &     4.30 &  \nodata &  \nodata \\
\ion{Fe}{1} & 4531.15 & 1.48 & $-$2.10 &    76.04 &     4.38 &  \nodata &  \nodata &    56.12 &     4.36 &  \nodata &  \nodata &    52.77 &     4.21 &    74.67 &     4.31 \\
\ion{Fe}{1} & 4592.65 & 1.56 & $-$2.46 &    73.61 &     4.79 &  \nodata &  \nodata &    47.13 &     4.65 &  \nodata &  \nodata &    35.04 &     4.34 &    49.72 &     4.35 \\
\ion{Fe}{1} & 4602.00 & 1.61 & $-$3.13 &  \nodata &  \nodata &  \nodata &  \nodata &  \nodata &  \nodata &  \nodata &  \nodata &  \nodata &  \nodata &    14.25 &     4.34 \\
\ion{Fe}{1} & 4602.94 & 1.49 & $-$2.21 &    81.57 &     4.59 &  \nodata &  \nodata &    55.64 &     4.46 &    73.38 &     4.47 &    57.03 &     4.39 &  \nodata &  \nodata \\
\ion{Fe}{1} & 4647.43 & 2.95 & $-$1.35 &    21.91 &     4.49 &  \nodata &  \nodata &  \nodata &  \nodata &  \nodata &  \nodata &  \nodata &  \nodata &  \nodata &  \nodata \\
\ion{Fe}{1} & 4733.59 & 1.49 & $-$2.99 &    46.97 &     4.78 &  \nodata &  \nodata &  \nodata &  \nodata &    32.74 &     4.53 &  \nodata &  \nodata &    31.61 &     4.47 \\
\ion{Fe}{1} & 4736.77 & 3.21 & $-$0.67 &    30.87 &     4.32 &  \nodata &  \nodata &  \nodata &  \nodata &    39.52 &     4.47 &    25.14 &     4.33 &    38.36 &     4.42 \\
\ion{Fe}{1} & 4871.32 & 2.87 & $-$0.34 &    71.95 &     4.24 &  \nodata &  \nodata &    67.68 &     4.43 &    58.21 &     4.02 &    53.78 &     4.11 &    59.56 &     4.01 \\
\ion{Fe}{1} & 4872.14 & 2.88 & $-$0.57 &    58.17 &     4.26 &  \nodata &  \nodata &    42.85 &     4.23 &  \nodata &  \nodata &    35.79 &     4.04 &    44.27 &     4.00 \\
\ion{Fe}{1} & 4890.76 & 2.88 & $-$0.38 &    65.90 &     4.19 &  \nodata &  \nodata &    54.13 &     4.24 &    82.57 &     4.50 &    52.79 &     4.15 &    63.93 &     4.13 \\
\ion{Fe}{1} & 4891.49 & 2.85 & $-$0.11 &    88.53 &     4.28 &  \nodata &  \nodata &    70.19 &     4.22 &    76.92 &     4.09 &    76.19 &     4.26 &    68.55 &     3.89 \\
\ion{Fe}{1} & 4903.31 & 2.88 & $-$0.89 &    43.72 &     4.35 &  \nodata &  \nodata &    32.14 &     4.36 &    39.71 &     4.28 &    36.79 &     4.38 &    33.73 &     4.14 \\
\ion{Fe}{1} & 4918.99 & 2.86 & $-$0.34 &    66.79 &     4.14 &  \nodata &  \nodata &    57.73 &     4.23 &  \nodata &  \nodata &  \nodata &  \nodata &  \nodata &  \nodata \\
\ion{Fe}{1} & 4920.50 & 2.83 &    0.07 &   100.27 &     4.29 &    29.72 &     3.52 &    87.19 &     4.34 &   101.02 &     4.34 &    71.06 &     3.96 &    88.68 &     4.04 \\
\ion{Fe}{1} & 4924.77 & 2.28 & $-$2.11 &    28.08 &     4.54 &  \nodata &  \nodata &  \nodata &  \nodata &    32.10 &     4.61 &  \nodata &  \nodata &    15.54 &     4.18 \\
\ion{Fe}{1} & 4938.81 & 2.88 & $-$1.08 &    30.39 &     4.30 &  \nodata &  \nodata &    26.24 &     4.42 &  \nodata &  \nodata &    22.37 &     4.26 &  \nodata &  \nodata \\
\ion{Fe}{1} & 4939.69 & 0.86 & $-$3.25 &    67.01 &     4.53 &  \nodata &  \nodata &    34.99 &     4.34 &  \nodata &  \nodata &    25.49 &     4.05 &    57.51 &     4.33 \\
\ion{Fe}{1} & 4966.09 & 3.33 & $-$0.79 &    26.31 &     4.48 &  \nodata &  \nodata &  \nodata &  \nodata &  \nodata &  \nodata &  \nodata &  \nodata &  \nodata &  \nodata \\
\ion{Fe}{1} & 4994.13 & 0.92 & $-$2.97 &    89.83 &     4.70 &  \nodata &  \nodata &    51.80 &     4.42 &  \nodata &  \nodata &    33.00 &     4.00 &  \nodata &  \nodata \\
\ion{Fe}{1} & 5005.71 & 3.88 & $-$0.12 &  \nodata &  \nodata &  \nodata &  \nodata &    21.22 &     4.51 &    33.79 &     4.62 &  \nodata &  \nodata &  \nodata &  \nodata \\
\ion{Fe}{1} & 5006.12 & 2.83 & $-$0.62 &    53.92 &     4.17 &  \nodata &  \nodata &  \nodata &  \nodata &  \nodata &  \nodata &  \nodata &  \nodata &    50.87 &     4.09 \\
\ion{Fe}{1} & 5049.82 & 2.28 & $-$1.36 &    54.10 &     4.22 &  \nodata &  \nodata &    40.50 &     4.26 &  \nodata &  \nodata &  \nodata &  \nodata &    56.64 &     4.22 \\
\ion{Fe}{1} & 5051.63 & 0.92 & $-$2.76 &  \nodata &  \nodata &  \nodata &  \nodata &  \nodata &  \nodata &    89.08 &     4.51 &    66.00 &     4.35 &  \nodata &  \nodata \\
\ion{Fe}{1} & 5079.22 & 2.20 & $-$2.10 &    46.36 &     4.74 &  \nodata &  \nodata &  \nodata &  \nodata &  \nodata &  \nodata &  \nodata &  \nodata &  \nodata &  \nodata \\
\ion{Fe}{1} & 5079.74 & 0.99 & $-$3.24 &  \nodata &  \nodata &  \nodata &  \nodata &  \nodata &  \nodata &    47.68 &     4.37 &  \nodata &  \nodata &  \nodata &  \nodata \\
\ion{Fe}{1} & 5083.34 & 0.96 & $-$2.84 &    91.64 &     4.64 &  \nodata &  \nodata &    45.19 &     4.22 &  \nodata &  \nodata &    56.16 &     4.30 &    85.98 &     4.49 \\
\ion{Fe}{1} & 5123.72 & 1.01 & $-$3.06 &    79.87 &     4.71 &  \nodata &  \nodata &  \nodata &  \nodata &    68.89 &     4.55 &  \nodata &  \nodata &    59.59 &     4.34 \\
\ion{Fe}{1} & 5127.36 & 0.92 & $-$3.25 &    76.01 &     4.72 &  \nodata &  \nodata &  \nodata &  \nodata &    51.98 &     4.36 &  \nodata &  \nodata &    42.73 &     4.16 \\
\ion{Fe}{1} & 5133.69 & 4.18 &    0.36 &    30.01 &     4.44 &  \nodata &  \nodata &  \nodata &  \nodata &  \nodata &  \nodata &  \nodata &  \nodata &  \nodata &  \nodata \\
\ion{Fe}{1} & 5150.84 & 0.99 & $-$3.04 &  \nodata &  \nodata &  \nodata &  \nodata &    47.81 &     4.50 &  \nodata &  \nodata &  \nodata &  \nodata &    59.23 &     4.29 \\
\ion{Fe}{1} & 5151.91 & 1.01 & $-$3.32 &    42.52 &     4.39 &  \nodata &  \nodata &  \nodata &  \nodata &    54.69 &     4.58 &  \nodata &  \nodata &  \nodata &  \nodata \\
\ion{Fe}{1} & 5166.28 & 0.00 & $-$4.12 &  \nodata &  \nodata &  \nodata &  \nodata &    57.57 &     4.53 &    76.21 &     4.43 &  \nodata &  \nodata &  \nodata &  \nodata \\
\ion{Fe}{1} & 5171.60 & 1.49 & $-$1.72 &   108.48 &     4.50 &  \nodata &  \nodata &    80.98 &     4.38 &  \nodata &  \nodata &    72.14 &     4.11 &   106.44 &     4.43 \\
\ion{Fe}{1} & 5191.45 & 3.04 & $-$0.55 &  \nodata &  \nodata &  \nodata &  \nodata &    40.09 &     4.33 &    62.29 &     4.48 &  \nodata &  \nodata &  \nodata &  \nodata \\
\ion{Fe}{1} & 5192.34 & 3.00 & $-$0.42 &    73.65 &     4.47 &  \nodata &  \nodata &    47.15 &     4.28 &    54.79 &     4.18 &  \nodata &  \nodata &    55.56 &     4.16 \\
\ion{Fe}{1} & 5194.94 & 1.56 & $-$2.02 &    82.81 &     4.41 &  \nodata &  \nodata &  \nodata &  \nodata &    61.58 &     4.08 &    62.02 &     4.31 &    78.41 &     4.29 \\
\ion{Fe}{1} & 5198.71 & 2.22 & $-$2.09 &    31.90 &     4.50 &  \nodata &  \nodata &  \nodata &  \nodata &  \nodata &  \nodata &  \nodata &  \nodata &  \nodata &  \nodata \\
\ion{Fe}{1} & 5202.34 & 2.18 & $-$1.87 &    65.77 &     4.77 &  \nodata &  \nodata &  \nodata &  \nodata &  \nodata &  \nodata &  \nodata &  \nodata &  \nodata &  \nodata \\
\ion{Fe}{1} & 5216.27 & 1.61 & $-$2.08 &    97.14 &     4.79 &  \nodata &  \nodata &    64.14 &     4.56 &    62.98 &     4.22 &    58.73 &     4.38 &    54.65 &     4.04 \\
\ion{Fe}{1} & 5225.53 & 0.11 & $-$4.76 &    50.99 &     4.81 &  \nodata &  \nodata &  \nodata &  \nodata &  \nodata &  \nodata &  \nodata &  \nodata &  \nodata &  \nodata \\
\ion{Fe}{1} & 5232.94 & 2.94 & $-$0.06 &    75.62 &     4.07 &  \nodata &  \nodata &    88.40 &     4.59 &    77.00 &     4.11 &  \nodata &  \nodata &  \nodata &  \nodata \\
\ion{Fe}{1} & 5263.31 & 3.27 & $-$0.87 &    21.87 &     4.37 &  \nodata &  \nodata &  \nodata &  \nodata &  \nodata &  \nodata &  \nodata &  \nodata &  \nodata &  \nodata \\
\ion{Fe}{1} & 5266.56 & 3.00 & $-$0.38 &    73.13 &     4.42 &  \nodata &  \nodata &    57.92 &     4.42 &  \nodata &  \nodata &    47.37 &     4.17 &    65.83 &     4.27 \\
\ion{Fe}{1} & 5281.79 & 3.04 & $-$0.83 &    56.44 &     4.66 &  \nodata &  \nodata &    29.36 &     4.41 &  \nodata &  \nodata &  \nodata &  \nodata &    37.60 &     4.32 \\
\ion{Fe}{1} & 5283.62 & 3.24 & $-$0.45 &    49.62 &     4.42 &  \nodata &  \nodata &    45.70 &     4.56 &  \nodata &  \nodata &  \nodata &  \nodata &  \nodata &  \nodata \\
\ion{Fe}{1} & 5307.36 & 1.61 & $-$2.91 &    27.14 &     4.46 &  \nodata &  \nodata &  \nodata &  \nodata &  \nodata &  \nodata &  \nodata &  \nodata &    13.73 &     4.05 \\
\ion{Fe}{1} & 5324.18 & 3.21 & $-$0.11 &    54.99 &     4.13 &  \nodata &  \nodata &    49.17 &     4.25 &    59.99 &     4.20 &    40.50 &     4.03 &    44.51 &     3.92 \\
\ion{Fe}{1} & 5332.90 & 1.56 & $-$2.78 &    41.41 &     4.52 &  \nodata &  \nodata &  \nodata &  \nodata &    31.20 &     4.33 &  \nodata &  \nodata &  \nodata &  \nodata \\
\ion{Fe}{1} & 5339.93 & 3.27 & $-$0.63 &    34.34 &     4.38 &  \nodata &  \nodata &  \nodata &  \nodata &  \nodata &  \nodata &  \nodata &  \nodata &    26.01 &     4.18 \\
\ion{Fe}{1} & 5341.02 & 1.61 & $-$1.95 &    92.77 &     4.56 &  \nodata &  \nodata &    55.28 &     4.27 &    95.88 &     4.66 &    52.27 &     4.13 &    85.55 &     4.39 \\
\ion{Fe}{1} & 5371.49 & 0.96 & $-$1.64 &   162.85 &     4.74 &    62.09 &     3.66 &   127.55 &     4.66 &   148.52 &     4.62 &  \nodata &  \nodata &   153.45 &     4.57 \\
\ion{Fe}{1} & 5383.37 & 4.31 &    0.64 &    35.43 &     4.41 &  \nodata &  \nodata &  \nodata &  \nodata &  \nodata &  \nodata &  \nodata &  \nodata &    24.96 &     4.17 \\
\ion{Fe}{1} & 5397.13 & 0.92 & $-$1.98 &   145.29 &     4.69 &  \nodata &  \nodata &   111.05 &     4.57 &   141.10 &     4.75 &   105.66 &     4.33 &   140.82 &     4.60 \\
\ion{Fe}{1} & 5405.77 & 0.99 & $-$1.85 &   152.72 &     4.79 &  \nodata &  \nodata &   110.77 &     4.52 &   115.14 &     4.15 &   112.04 &     4.43 &   138.98 &     4.52 \\
\ion{Fe}{1} & 5410.91 & 4.47 &    0.40 &    20.97 &     4.53 &  \nodata &  \nodata &  \nodata &  \nodata &  \nodata &  \nodata &  \nodata &  \nodata &  \nodata &  \nodata \\
\ion{Fe}{1} & 5415.20 & 4.39 &    0.64 &    29.97 &     4.39 &  \nodata &  \nodata &  \nodata &  \nodata &  \nodata &  \nodata &  \nodata &  \nodata &    27.94 &     4.33 \\
\ion{Fe}{1} & 5429.70 & 0.96 & $-$1.88 &   156.21 &     4.84 &    50.60 &     3.68 &  \nodata &  \nodata &   124.83 &     4.35 &   115.17 &     4.50 &   135.04 &     4.42 \\
\ion{Fe}{1} & 5434.52 & 1.01 & $-$2.13 &   135.97 &     4.77 &  \nodata &  \nodata &    99.79 &     4.57 &   125.76 &     4.68 &  \nodata &  \nodata &   125.87 &     4.54 \\
\ion{Fe}{1} & 5497.52 & 1.01 & $-$2.82 &   103.78 &     4.82 &  \nodata &  \nodata &    53.69 &     4.37 &    72.68 &     4.33 &  \nodata &  \nodata &    87.33 &     4.49 \\
\ion{Fe}{1} & 5501.47 & 0.96 & $-$3.05 &    88.68 &     4.73 &  \nodata &  \nodata &  \nodata &  \nodata &    73.55 &     4.51 &    46.93 &     4.33 &    75.21 &     4.46 \\
\ion{Fe}{1} & 5506.78 & 0.99 & $-$2.79 &   102.20 &     4.74 &  \nodata &  \nodata &    73.82 &     4.66 &    81.74 &     4.42 &    53.96 &     4.22 &    80.17 &     4.32 \\
\ion{Fe}{1} & 5569.62 & 3.42 & $-$0.52 &    38.06 &     4.51 &  \nodata &  \nodata &  \nodata &  \nodata &  \nodata &  \nodata &  \nodata &  \nodata &  \nodata &  \nodata \\
\ion{Fe}{1} & 5586.76 & 3.37 & $-$0.11 &    54.31 &     4.29 &  \nodata &  \nodata &  \nodata &  \nodata &    49.85 &     4.22 &    39.23 &     4.19 &    48.84 &     4.17 \\
\ion{Fe}{1} & 5624.54 & 3.42 & $-$0.76 &    19.58 &     4.36 &  \nodata &  \nodata &  \nodata &  \nodata &  \nodata &  \nodata &  \nodata &  \nodata &  \nodata &  \nodata \\
\ion{Fe}{1} & 5701.54 & 2.56 & $-$2.14 &    14.91 &     4.54 &  \nodata &  \nodata &  \nodata &  \nodata &  \nodata &  \nodata &  \nodata &  \nodata &  \nodata &  \nodata \\
\ion{Fe}{1} & 6065.48 & 2.61 & $-$1.41 &    36.02 &     4.32 &  \nodata &  \nodata &    29.36 &     4.43 &    25.51 &     4.11 &  \nodata &  \nodata &    25.16 &     4.07 \\
\ion{Fe}{1} & 6136.61 & 2.45 & $-$1.41 &    55.69 &     4.42 &  \nodata &  \nodata &    29.32 &     4.23 &    68.42 &     4.62 &    41.18 &     4.38 &    49.45 &     4.28 \\
\ion{Fe}{1} & 6137.69 & 2.59 & $-$1.35 &    54.51 &     4.52 &  \nodata &  \nodata &    26.98 &     4.29 &    33.77 &     4.18 &  \nodata &  \nodata &    34.20 &     4.16 \\
\ion{Fe}{1} & 6191.56 & 2.43 & $-$1.42 &    54.46 &     4.38 &  \nodata &  \nodata &    51.81 &     4.61 &    66.73 &     4.57 &    37.23 &     4.29 &    41.41 &     4.14 \\
\ion{Fe}{1} & 6219.28 & 2.20 & $-$2.45 &    20.01 &     4.51 &  \nodata &  \nodata &  \nodata &  \nodata &  \nodata &  \nodata &  \nodata &  \nodata &  \nodata &  \nodata \\
\ion{Fe}{1} & 6230.72 & 2.56 & $-$1.28 &    60.40 &     4.49 &  \nodata &  \nodata &    47.51 &     4.56 &    35.51 &     4.10 &    37.90 &     4.32 &    42.18 &     4.17 \\
\ion{Fe}{1} & 6246.32 & 3.60 & $-$0.77 &  \nodata &  \nodata &  \nodata &  \nodata &  \nodata &  \nodata &    25.66 &     4.70 &  \nodata &  \nodata &  \nodata &  \nodata \\
\ion{Fe}{1} & 6252.56 & 2.40 & $-$1.77 &    47.19 &     4.58 &  \nodata &  \nodata &  \nodata &  \nodata &  \nodata &  \nodata &  \nodata &  \nodata &    33.97 &     4.33 \\
\ion{Fe}{1} & 6265.13 & 2.18 & $-$2.54 &  \nodata &  \nodata &  \nodata &  \nodata &  \nodata &  \nodata &  \nodata &  \nodata &  \nodata &  \nodata &    16.09 &     4.41 \\
\ion{Fe}{1} & 6393.60 & 2.43 & $-$1.58 &    56.72 &     4.56 &  \nodata &  \nodata &    37.93 &     4.52 &    29.29 &     4.12 &  \nodata &  \nodata &    32.72 &     4.14 \\
\ion{Fe}{1} & 6421.35 & 2.28 & $-$2.01 &    40.98 &     4.57 &  \nodata &  \nodata &  \nodata &  \nodata &    33.33 &     4.43 &    19.49 &     4.31 &    28.60 &     4.30 \\
\ion{Fe}{1} & 6430.85 & 2.18 & $-$1.95 &    44.82 &     4.44 &  \nodata &  \nodata &    30.78 &     4.46 &    36.20 &     4.29 &  \nodata &  \nodata &    32.50 &     4.19 \\
\ion{Fe}{1} & 6494.98 & 2.40 & $-$1.24 &    71.82 &     4.39 &  \nodata &  \nodata &    44.70 &     4.26 &    60.55 &     4.24 &    39.84 &     4.10 &    59.53 &     4.17 \\
\ion{Fe}{1} & 6592.91 & 2.73 & $-$1.47 &  \nodata &  \nodata &  \nodata &  \nodata &  \nodata &  \nodata &  \nodata &  \nodata &  \nodata &  \nodata &    19.87 &     4.12 \\
\ion{Fe}{1} & 6677.99 & 2.69 & $-$1.42 &    46.40 &     4.56 &  \nodata &  \nodata &    31.25 &     4.54 &    24.53 &     4.16 &  \nodata &  \nodata &    24.91 &     4.14 \\
\ion{Fe}{1} & 7495.07 & 4.22 & $-$0.10 &    23.83 &     4.69 &  \nodata &  \nodata &  \nodata &  \nodata &  \nodata &  \nodata &  \nodata &  \nodata &  \nodata &  \nodata \\
\ion{Fe}{1} & 7511.02 & 4.18 &    0.12 &  \nodata &  \nodata &  \nodata &  \nodata &  \nodata &  \nodata &  \nodata &  \nodata &  \nodata &  \nodata &    16.52 &     4.20 \\
\ion{Fe}{2} & 4173.45 & 2.58 & $-$2.38 &  \nodata &  \nodata &  \nodata &  \nodata &  \nodata &  \nodata &  \nodata &  \nodata &    38.79 &     4.27 &  \nodata &  \nodata \\
\ion{Fe}{2} & 4178.86 & 2.58 & $-$2.51 &  \nodata &  \nodata &  \nodata &  \nodata &  \nodata &  \nodata &    41.48 &     4.34 &  \nodata &  \nodata &    36.60 &     4.15 \\
\ion{Fe}{2} & 4233.16 & 2.58 & $-$2.02 &    84.87 &     4.52 &  \nodata &  \nodata &  \nodata &  \nodata &  \nodata &  \nodata &    67.79 &     4.44 &  \nodata &  \nodata \\
\ion{Fe}{2} & 4385.38 & 2.78 & $-$2.64 &    34.29 &     4.47 &  \nodata &  \nodata &  \nodata &  \nodata &    34.80 &     4.57 &  \nodata &  \nodata &    20.20 &     4.16 \\
\ion{Fe}{2} & 4416.82 & 2.78 & $-$2.57 &    33.91 &     4.39 &  \nodata &  \nodata &    26.25 &     4.51 &    22.71 &     4.24 &  \nodata &  \nodata &    32.17 &     4.36 \\
\ion{Fe}{2} & 4491.41 & 2.86 & $-$2.71 &    28.85 &     4.52 &  \nodata &  \nodata &  \nodata &  \nodata &  \nodata &  \nodata &  \nodata &  \nodata &  \nodata &  \nodata \\
\ion{Fe}{2} & 4508.28 & 2.86 & $-$2.42 &  \nodata &  \nodata &  \nodata &  \nodata &    30.67 &     4.55 &  \nodata &  \nodata &  \nodata &  \nodata &    35.11 &     4.36 \\
\ion{Fe}{2} & 4515.34 & 2.84 & $-$2.60 &    34.32 &     4.50 &  \nodata &  \nodata &    23.99 &     4.55 &  \nodata &  \nodata &    25.20 &     4.50 &    22.84 &     4.25 \\
\ion{Fe}{2} & 4555.89 & 2.83 & $-$2.40 &    42.46 &     4.43 &  \nodata &  \nodata &    31.69 &     4.51 &    29.75 &     4.28 &    22.40 &     4.21 &  \nodata &  \nodata \\
\ion{Fe}{2} & 4583.83 & 2.81 & $-$1.94 &    80.86 &     4.58 &  \nodata &  \nodata &  \nodata &  \nodata &    71.08 &     4.53 &    49.30 &     4.26 &    58.96 &     4.22 \\
\ion{Fe}{2} & 5197.58 & 3.23 & $-$2.22 &    27.15 &     4.41 &  \nodata &  \nodata &  \nodata &  \nodata &  \nodata &  \nodata &  \nodata &  \nodata &  \nodata &  \nodata \\
\ion{Fe}{2} & 5276.00 & 3.20 & $-$2.01 &  \nodata &  \nodata &  \nodata &  \nodata &    34.24 &     4.56 &  \nodata &  \nodata &  \nodata &  \nodata &    32.31 &     4.27 \\
\ion{Co}{1} & 3845.47 & 0.92 &    0.06 &  \nodata &  \nodata &  \nodata &  \nodata &  \nodata &  \nodata &  \nodata &  \nodata &  \nodata &  \nodata &    82.64 &     1.61 \\
\ion{Co}{1} & 3894.08 & 1.05 &    0.12 &  \nodata &  \nodata &  \nodata &  \nodata &  \nodata &  \nodata &  \nodata &  \nodata &  \nodata &  \nodata &    77.72 &     1.59 \\
\ion{Co}{1} & 3995.31 & 0.92 & $-$0.18 &    47.73 &     1.19 &  \nodata &  \nodata &    68.76 &     1.95 &    76.87 &     1.76 &    63.18 &     1.73 &    68.56 &     1.51 \\
\ion{Co}{1} & 4121.32 & 0.92 & $-$0.33 &  \nodata &  \nodata &  \nodata &  \nodata &  \nodata &  \nodata &    55.80 &     1.48 &  \nodata &  \nodata &  \nodata &  \nodata \\
\ion{Ni}{1} & 3783.53 & 0.42 & $-$1.40 &  \nodata &  \nodata &  \nodata &  \nodata &  \nodata &  \nodata &  \nodata &  \nodata &    78.78 &     2.61 &  \nodata &  \nodata \\
\ion{Ni}{1} & 3807.14 & 0.42 & $-$1.23 &  \nodata &  \nodata &    73.92 &     2.83 &  \nodata &  \nodata &  \nodata &  \nodata &  \nodata &  \nodata &  \nodata &  \nodata \\
\ion{Ni}{1} & 4714.42 & 3.38 &    0.25 &    12.77 &     2.85 &  \nodata &  \nodata &    20.09 &     3.25 &  \nodata &  \nodata &  \nodata &  \nodata &  \nodata &  \nodata \\
\ion{Ni}{1} & 5476.90 & 1.83 & $-$0.78 &    70.64 &     3.01 &  \nodata &  \nodata &  \nodata &  \nodata &    55.49 &     2.79 &  \nodata &  \nodata &    61.07 &     2.82 \\
\ion{Zn}{1} & 4722.16 &    4.03  &  $-$0.37  &     syn &     1.40  & \nodata &  \nodata & \nodata &  \nodata & \nodata &  \nodata & \nodata &  \nodata &     syn &     1.51  \\
\ion{Zn}{1} & 4810.54 &    4.08  &  $-$0.15  &     syn &     1.36  & \nodata &  \nodata &     syn &     2.06 &     syn &     1.76 & \nodata &  \nodata & \nodata &  \nodata  \\
\ion{Sr}{2} & 4077.71 &    0.00  &     0.15  &     syn &  $-$1.64  &     syn &  $-$1.33 &     syn &  $-$0.20 &     syn &  $-$0.48 &     syn &  $-$0.37 &     syn &  $-$0.82  \\
\ion{Sr}{2} & 4215.52 &    0.00  &  $-$0.17  &     syn &  $-$1.56  &     syn &  $-$1.27 &     syn &  $-$0.22 &     syn &  $-$0.44 &     syn &  $-$0.37 &     syn &  $-$1.01  \\
\ion{Ba}{2} & 4554.03 &    0.00  &     0.16  &     syn &  $-$2.46  &     syn &  $-$2.64 &     syn &  $-$1.74 &     syn &  $-$1.47 &     syn &  $-$1.56 &     syn &  $-$1.54  \\
\ion{Ba}{2} & 4934.09 &    0.00  &  $-$0.16  &     syn &  $-$2.43  &     syn &  $-$2.69 &     syn &  $-$1.72 &     syn &  $-$1.57 &     syn &  $-$1.46 &     syn &  $-$1.54  \\
\ion{Ba}{2} & 6141.71 &    0.70  &  $-$0.01  &     syn &  $-$2.38  & \nodata &  \nodata &     syn &  $-$1.67 &     syn &  $-$1.52 &     syn &  $-$1.56 &     syn &  $-$1.58  \\
\ion{Ba}{2} & 6496.90 &    0.60  &  $-$0.37  & \nodata &  \nodata  & \nodata &  \nodata &     syn &  $-$1.67 &     syn &  $-$1.57 &     syn &  $-$1.31 &     syn &  $-$1.49  \\

\enddata
\end{deluxetable}


\begin{deluxetable*}{lrrrrrrrl}[!ht]
\tabletypesize{\small}
\tabletypesize{\footnotesize}
\tablewidth{0pc}
\tablecaption{NLTE corrections for the program stars \label{nlte}}
\tablehead{
\colhead{Ion}         &
\colhead{\jone\vv}    &
\colhead{\jtwo\xx}    &
\colhead{\jthree}     &
\colhead{\jfour}      &
\colhead{\jfive}      &
\colhead{\jsix\yy}    &
\colhead{$N$}         &
\colhead{Reference}   }
\startdata
\ion{Na}{1}   & $-$0.26 &    0.02 & $-$0.22 & $-$0.41 & $-$0.26 & $-$0.27 &   2 & \citet{lind2011}        \\
\ion{Mg}{1}   &    0.11 &    0.22 &    0.07 &    0.07 &    0.08 &    0.13 &   5 & \citet{bergemann2015}   \\
\ion{Al}{1}   &    1.00 &    0.70 &    1.00 &    1.00 &    1.00 &    1.00 &   1 & \citet{nordlander2017b} \\
\ion{Si}{1}   & $-$0.02 & \nodata & $-$0.02 & $-$0.04 & $-$0.03 & $-$0.04 &   2 & \citet{bergemann2013}   \\
\ion{Ca}{1}   &    0.24 &    0.28 &    0.23 &    0.25 &    0.25 &    0.24 &   9 & \citet{mashonkina2007}  \\
\ion{Ti}{1}   &    0.54 &    0.65 &    0.58 &    0.57 &    0.59 &    0.55 &   7 & \citet{bergemann2011}   \\
\ion{Ti}{2}   &    0.09 &    0.07 &    0.08 &    0.11 &    0.10 &    0.11 &  22 & \citet{bergemann2011}   \\
\ion{Cr}{1}   &    0.54 &    0.87 &    0.55 &    0.53 &    0.57 &    0.59 &   6 & \citet{bergemann2010}   \\
\ion{Mn}{1}   &    0.36 & \nodata &    0.48 &    0.33 &    0.48 &    0.64 &   2 & \citet{bergemann2019}   \\
\ion{Fe}{1}   &    0.19 &    0.30 &    0.13 &    0.13 &    0.13 &    0.12 & 148 & \citet{bergemann2012b}  \\
\ion{Fe}{2}   &    0.00 & \nodata &    0.00 &    0.00 &    0.00 &    0.00 &  12 & \citet{bergemann2012b}  \\
\ion{Co}{1}   &    0.79 & \nodata &    0.74 &    0.79 &    0.76 &    0.84 &   4 & \citet{bergemann2010b}  \\
\enddata
\tablenotetext{a}{Assumed \logg=1.0 for \ion{Na}{1} calculations.}
\tablenotetext{b}{Assumed \metal=$-4.0$ for \ion{Na}{1} calculations.}
\tablenotetext{c}{Assumed \logg=1.0 for \ion{Na}{1} calculations.}
\end{deluxetable*}

\end{document}